\documentclass[aps,pre,floats,floatfix,superscriptaddress]{revtex4}
\usepackage{latexsym,amsmath}
\usepackage{amsbsy}
\usepackage[usenames]{color}
\usepackage[pdftex]{graphicx}
\usepackage{amsfonts,amsmath,amssymb,hyperref}

\usepackage{color}
\usepackage{todonotes}
\usepackage{algorithm}
\usepackage{algorithmic}
\usepackage[usenames]{color}
\usepackage{bbm}
\usepackage{array}
\usepackage{multirow}
\usepackage{makecell}

\begin{document}
\title{Complementarity in Complex Networks}

\author{Gabriel Budel}
\affiliation{Delft University of Technology, Faculty of Electrical Engineering, Mathematics and Computer Science, 2628 CD, Delft, The Netherlands}
\author{Maksim Kitsak}
\affiliation{Delft University of Technology, Faculty of Electrical Engineering, Mathematics and Computer Science, 2628 CD, Delft, The Netherlands}

\begin{abstract}
In many networks, including networks of protein-protein interactions, interdisciplinary collaboration networks, and semantic networks, connections are established between nodes with complementary rather than similar properties. While complementarity is abundant in networks, we lack mathematical intuition and quantitative methods to study complementarity mechanisms in these systems. In this work, we close this gap by providing a rigorous definition of complementarity and developing geometric complementarity frameworks for modeling and inference tasks on networks. We demonstrate the utility of complementarity frameworks by learning geometric representations of several real systems. Complementarity not only offers novel practical analysis methods but also enhances our intuition about formation mechanisms in networks on a broader scale and calls for a careful re-evaluation of existing similarity-inspired methods.
\end{abstract}

\maketitle

\section{Introduction: complementarity versus similarity in complex networks}

Complementarity plays an important role in many real networks, including networks of molecular interactions, scientific collaborations, semantic networks, and 
production networks. Similar protein molecules are not guaranteed to interact. Instead, interacting proteins seem to exhibit complementarity at several levels, including shape electrostatic complementarity~\cite{mccoy1997electrostatic}, hydrophobic mismatch~\cite{lewis1983lipid,yano2006measurement,botelho2006curvature}, and shape complementarity~\cite{zhang2009shape,li2013role}. Collaboration teams, be it scientific collaborators, advisory boards, or military units, greatly benefit from combining experts that complement each other's knowledge and skills. Semantic networks form another domain where complementarity could play an important role. Indeed, our language would be rather bleak if we uses only similar words in a sentence. Comparisons, generalizations, and analogies are only a few out of many examples of elements that make for a colorful representation in modern language. All these examples indicate that we construct sentences with words that are complementary rather than similar in meaning. Finally, a very recent work finds that complementarity plays a key role in the formation of production networks, since companies are especially similar to their close competitors but not to their trading partners~\cite{mattsson2021functional}.

{\it What is complementarity?} Intuitively, one object complements another one by contributing properties or features lacking in the other object. Unfortunately, our understanding of complementarity and its mechanisms in network formation and dynamics does not go far beyond this general definition. We lack both an intuitive understanding of complementarity and methods to quantify complementarity mechanisms in networks. In this work, we aim to fill in both gaps by proposing a principled complementarity framework and developing methods to learn complementarity representations of real networks.

Complementarity-based networks are routinely analyzed using similarity-inspired methods. This is the case since the mechanisms of similarity are much better understood. Indeed, similarity plays a key role in the formation of social interactions -- the more similar individuals are, the higher the chance of their interaction. Social interactions are relatively easy to measure and document, and, historically, the first studies of social networks can be traced back to the beginning of the 20th century. Much later, in the early 2000s, the availability of large-scale social networks fueled the rise of network and data sciences. As a result, the majority of existing network analysis and inference techniques are either developed for or inspired by social networks.

Another reason for the wide acceptance of similarity methods is, arguably, their geometric interpretation. Network nodes can be viewed as points in a certain latent spaces, such that similarities are functions of distances between the corresponding points. The smaller the distance between points, the larger the similarity between the nodes. These geometric frameworks can be traced back to sociology in the 1970s~\cite{mcfarland1973social}. It is the similarity interpretation of distances in the latent space that leads to a large array of network analysis methods, including link prediction~\cite{cannistraci2013minimum,yang2015embedding,xiao2015from,tang2015line,Grover2016node2vec,zhu2016scalable,nickel2018learning,Kazemi2018SimplE,brew2010latent,kitsak2020link,perez2020precision}, soft community detection and clustering~\cite{newman2015generalized,zuev2015emergence,Yang2015unified,Sewell2017latent}, network navigation~\cite{Boguna2010sustaining,Gulyas2015,voitalov2017geohyperbolic,Garcia2018multiscale}, and search~\cite{kleinberg2006complex,ratnasamy2001scalable,risson2006survey}.

\begin{figure}
\includegraphics[width=6.5in]{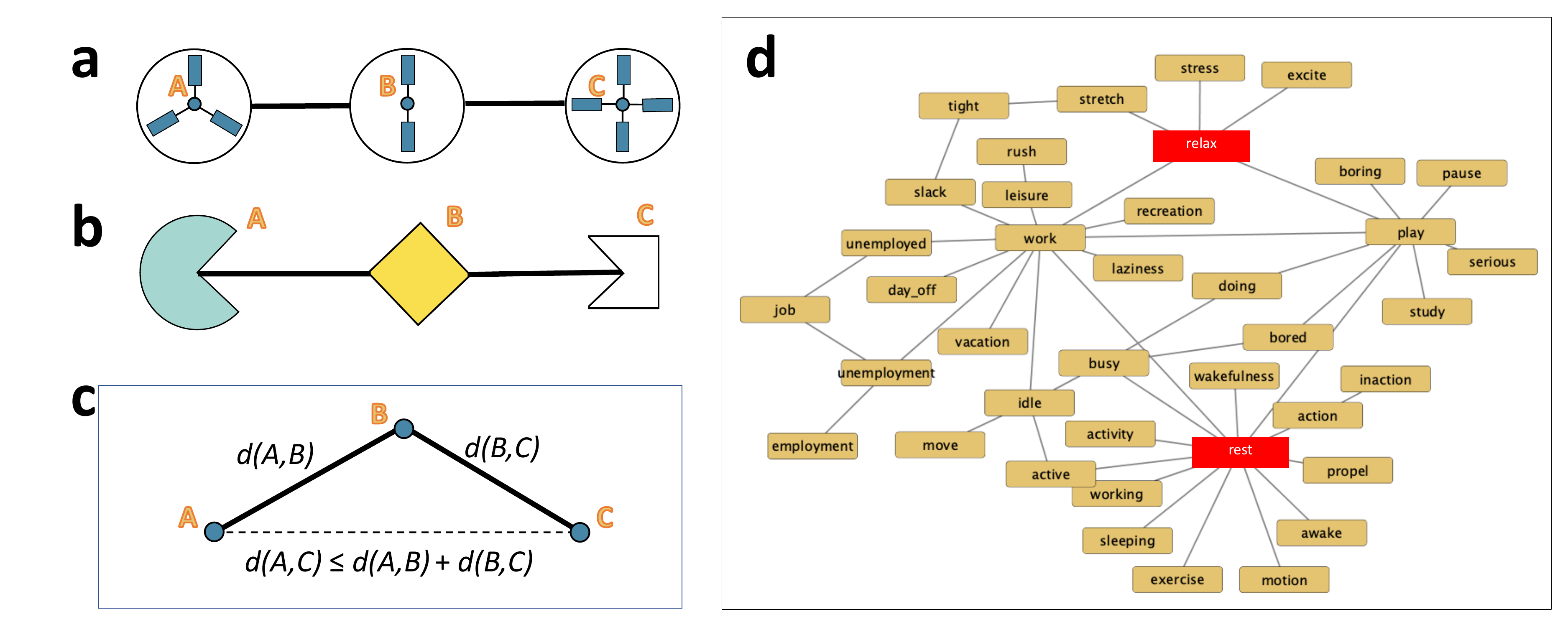}
\caption{\footnotesize (color online) ({\bf a}) Similarity-based toy network. Nodes are depicted as propellers with variable numbers of blades, all nodes are similar since they are all propellers. Similarity is transitive: if A is similar to B, and B is similar to C, then A is expected to be similar to C. ({\bf b}) Complementarity-based toy network: complementarity of two nodes is shown by matching interfaces. Complementarity is not transitive: if A is complementary to B (there are matching interfaces) and B is complementary to C (there are matching interfaces), A is not guaranteed to be complementary to C (there are no matching interfaces).
({\bf c})  Embedding of the toy network into a metric space imposes constraints on distances due to the triangle inequality. ({\bf d}) A subgraph of the antonym network. Note that the \emph{rest-relax} and \emph{unemployed-unemployment} node pairs share 2 common neighbors each and can be incorrectly interpreted as being connected if similarity-based methods are blindly applied to the network.}
\label{fig:1}
\end{figure}

The success of network embeddings in representations of social networks -- where node similarity is recognized as one of the key mechanisms for link formation -- can be attributed to the agreement between the transitive property of similarity and the metric property of the latent spaces used in network embeddings. Indeed, consider the similarity-shaped toy network in Fig.~\ref{fig:1}{\bf a}. If node $A$ is similar to node $B$ and node $B$ is similar to $C$, by transitivity, $A$ is expected to be similar to $C$. A generic network embedding, Fig.~\ref{fig:1}{\bf c}, would result in a small distance $d(A,B)$ since $A$ and $B$ are connected. Likewise, $d(B,C)$ is also expected to be small since nodes $B$ and $C$ are connected. Then $d(A,C)$ must be small by the triangle inequality:  $d(A,C) \leq d(A,B) + d(B,C)$, implying that $A$ and $C$ are similar and may be connected. 

{\it Why are network embedding methods not readily applicable to complementarity-driven systems?} The answer is rooted in the non-transitivity of complementarity. Imagine that the same toy network is now formed by the principles of complementarity,  Fig.~\ref{fig:1}{\bf b}. The complementarity of $A$ and $B$ and of $B$ and $C$ {\it does not} imply complementarity of $A$ and $C$. The complementarity network in Fig.~\ref{fig:1}{\bf b}, however, is identical to its similarity counterpart in Fig.~\ref{fig:1}{\bf a}, resulting in the same embedding as in Fig.~\ref{fig:1}{\bf c}. Consequently, a small distance $d(A,C)$ in the embedding of the complementarity network immediately translates to inference errors. If the goal of the network embedding is to predict missing links, for instance, a small distance $d(A,C)$ implies a false positive link candidate between $A$ and $C$. If, on the other hand, the problem at hand is cluster analysis, $A$, $B$, and $C$ may end up in the same cluster of nodes in $\mathcal{M}$, implying that they all have similar characteristics, which is not the case for a complementarity-based network. As a real example, we consider a subgraph of the antonym network, which can be regarded as a colloquial complementarity network. Here nodes are words and links are established between words with opposite meanings, Fig.~\ref{fig:1}{\bf d}.  The ``rest-relax" word pair are synonyms, in the antonym network these two words share two antonyms: ``work" and ``play". Being synonyms, the words ``rest" and ``relax" are not connected in the antonym network. If a {\it similarity}-based link prediction algorithm is blindly applied to the network, however, the ``rest - relax" pair can be interpreted as a missing link candidate, contrary to our intuition.

The demonstrated non-transitivity of complementarity principles invites us to revise existing network-based and embedding-based approaches to make them applicable to complementarity-driven networks, which we do in the following sections of the manuscript. In section II, we define the principled complementarity framework, quantifying complementarity between nodes as their ability to execute certain functions or tasks. In section III, we analyze the properties of the principled complementarity framework to deduce a minimal practical framework for learning representations of real systems. In section IV, we apply the minimal complementarity framework to learn complementarity representations of five real networks. We conclude the manuscript with the summary and outlook in Section V.

The idea that the complementarity principle is different from similarity is quickly getting traction in the scientific community. Therefore, before proceeding further, we find it important to pause and acknowledge other concurrent efforts to study complementarity mechanisms in networks. Shortly after the initial work of I. Kov\'{a}cs highlighting the role of complementarity in protein-protein interaction networks~\cite{kovacs2019network} and the first version of the present work, C. Mattson et al. discovered the prominent role of complementarity in production networks~\cite{mattsson2021functional}. Another very recent work by S. Talaga and A. Nowak proposes to quantify the relative role of similarity and complementarity in a network through densities of triangle and quadrangle subgraphs in it~\cite{talaga2022structural}. While these works are invaluable in developing our intuition about complementarity in real networks, we lack rigorous mathematical frameworks that would allow us to model synthetic complementarity networks and learn complementarity representations of real networks. We address both challenges in the following Sections.

\section{Towards Principled Complementarity framework}

\begin{figure}
\includegraphics[width=6.5in]{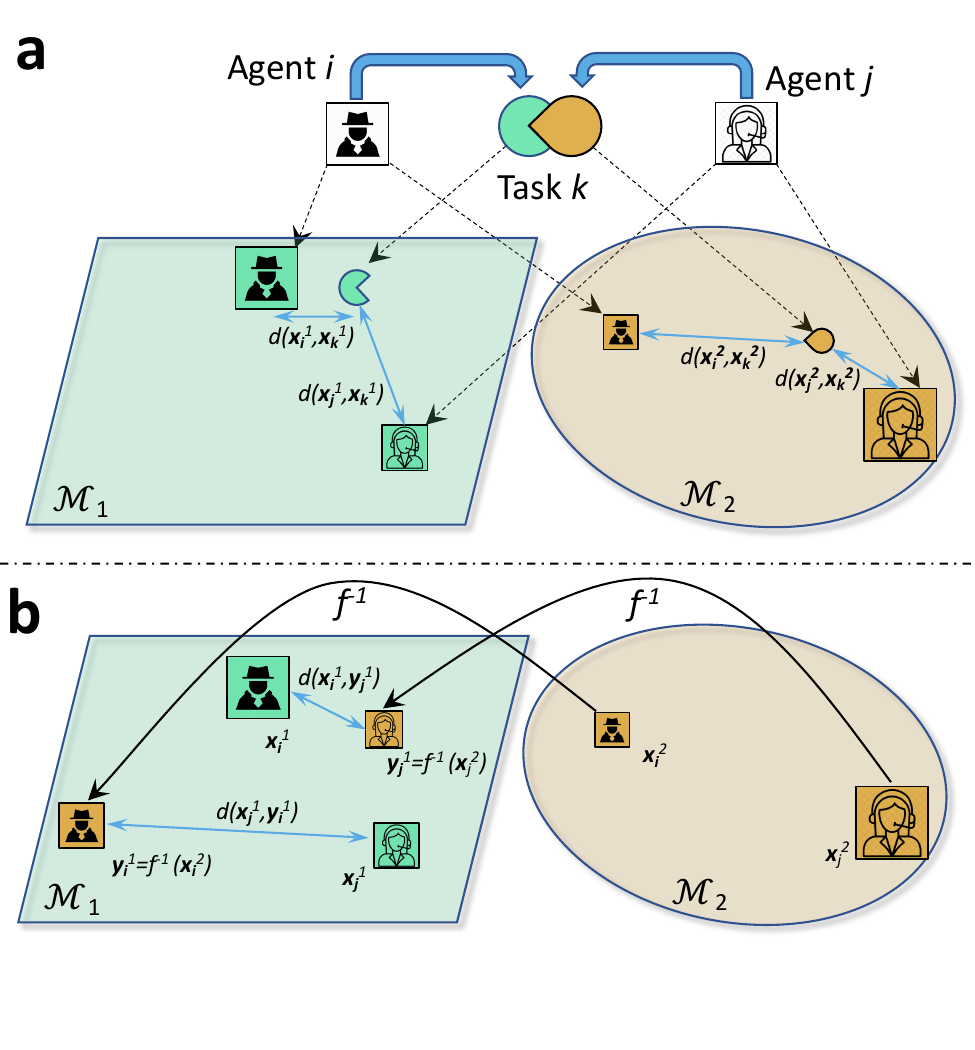}
\caption{\footnotesize (color online) {\bf a}, The principled complementarity framework. Agents $i$ and $j$ complement each other in solving interdisciplinary task $k$. The two disciplines related to task $k$ are represented as latent metric spaces $\mathcal{M}_{1}$ and $\mathcal{M}_{2}$.
Task $k$ consists of two independent disciplinary parts that are represented as points $\mathbf{x}_{k}^{1}$ and $\mathbf{x}_{k}^{2}$ in spaces $\mathcal{M}_{1}$ and $\mathcal{M}_{2}$, respectively. Likewise, the skill sets of agents $i$ and $j$ are points $\mathbf{x}_{i}^{1,2}$ and $\mathbf{x}_{j}^{1,2}$, respectively. Agent $i$ has a higher probability to solve subtask in $\mathcal{M}_{1}$ if $d\left(x_{i}^{1}, x_{k}^{1}\right) < d\left(x_{j}^{1}, x_{k}^{1}\right)$. Similarly, agent $j$ has a higher probability to solve subtask in $\mathcal{M}_{2}$ if $d\left(x_{j}^{2}, x_{k}^{2}\right) < d\left(x_{i}^{2}, x_{k}^{2}\right)$ .
{\bf b}, The minimal complementarity framework is a special case of the principled complementarity framework when task coordinates in $\mathcal{M}_{1}$ and $\mathcal{M}_{2}$ are related by an injective continuous map $f:\mathcal{M}_{1} \to \mathcal{M}_{2}$ for $k = 1,\ldots,T$. In this case, agent coordinates in $\mathcal{M}_{2}$ are effectively mapped to $\mathcal{M}_{1}$, $\mathbf{y}^{1}_{i} = f^{-1}\left(\mathbf{x}^{2}_{i} \right)$ for $i = 1,\ldots,N$, see Eq.~(\ref{eq:compl2}). As a result, the complementarity between agents $i$ and $j$ is quantified by distances $d(\mathbf{x}^{1}_{i}, \mathbf{y}^{1}_{j})$ and $d(\mathbf{x}^{1}_{j}, \mathbf{y}^{1}_{i})$, the smaller the distances the higher the complementarity.
}
\label{fig:compl}
\end{figure}

To define complementarity from \emph{first principles}, we refer to the Oxford English Dictionary, which asserts that \emph{``two people or things that are complementary are different but together form a useful or attractive combination of skills, qualities or physical features"}.

This definition implies that complementarity can be defined if each network node is characterized by at least two different skill types or features that might be able to complement each other. These features are intrinsic properties of network nodes that may or may not be readily observable. In the context of scientific collaborations, the features are the expertises of researchers in different disciplines. In the context of molecular interactions, the features might correspond to chemical properties of molecules of interest. 

To define complementarity between nodes $i$ and $j$ we need to quantify the extent to which $i$ and $j$ form a useful combination of features or skills. One way to define such usefulness is through tasks or functions that nodes $i$ and $j$ can jointly execute. Depending on the system of interest, the task can be either a scientific problem  in the case of a scientific collaboration network, or a biological function that two molecules can execute by forming a physical interaction in the case of a PPI network.  Tasks mediate complementarity between agents and are also characterized by two features each. 

All aspects considered, to introduce a complementarity-based network we introduce $N$ nodes, each of which is characterized by two skills  $\mathbf{x}_{i}^{1}$ and $\mathbf{x}_{i}^{2}$, $i = 1,\ldots, N$, and $T$ tasks that are also characterized by two skills each, $\mathbf{x}_{k}^{1}$ and $\mathbf{x}_{k}^{2}$, $k = 1, \ldots,T$. 

To quantify the complementarity between any two nodes $i$ and $j$, we need to be able to compare their skills. To do so, we postulate that skills are nothing else but points in two latent spaces $\mathcal{M}_{1}$ and $\mathcal{M}_{2}$. Therefore, distances can be defined between features of the same type but not necessarily between features of different types. In the case of scientific collaboration, for instance, it is straightforward to compare two skill sets within the same discipline but it is not for two skill sets from different disciplines. Thus, in agreement with the similarity intuition, we postulate that the distance $d(\mathbf{x}_{i}^{a}, \mathbf{x}_{j}^{a})$ between two nodes $i$ and $j$ in the space $\mathcal{M}_{a}$ quantifies the similarity between the features $\mathbf{x}_{i}^{a}$ and $\mathbf{x}_{j}^{a}$: the smaller the distance $d(\mathbf{x}_{i}^{a}, \mathbf{x}_{j}^{a})$, the larger the similarity between $i$ and $j$ with respect to skill $a$. 

We are now in a good position to define complementarity. The complementarity between any two nodes $i$ and $j$ with respect to task $k$ is the probability $p(i,j|k)$ that the two agents $i$ and $j$ can jointly execute $k$. In the most basic setting, one can think of task $k$ as consisting of two independent parts $1$ and $2$ corresponding to $\mathcal{M}_{1}$ and $\mathcal{M}_{2}$, respectively. Then, task $k$ can be executed either as a result of node $i$ executing part $1$ and node $j$ executing part $2$ or vice versa:
\begin{equation}
\label{eq:compl1}
p(i,j|k) = 1 - \left[1- r_{1}\left(d\left[\mathbf{x}_{i}^{1},\mathbf{x}_{k}^{1}\right]\right)r_{2}\left(d\left[\mathbf{x}_{j}^{2},\mathbf{x}_{k}^{2}\right]\right)\right]\left[1- r_{1}\left(d\left[\mathbf{x}_{j}^{1},\mathbf{x}_{k}^{1}\right]\right)r_{2}\left(d\left[\mathbf{x}_{i}^{2},\mathbf{x}_{k}^{2}\right]\right)\right].
\end{equation}

Here, $r_{a}\left(d\left[\mathbf{x}_{i}^{a},\mathbf{x}_{k}^{a}\right]\right)$ is the probability that node $i$ can independently execute part $a$ of task $k$, $a={1,2}$. Our physical intuition suggests that connection probabilities $r_{\{1,2\}}(d)$ are decreasing functions of distances $d$ since the smaller the distance, the more similar the agent's expertise $\mathbf{x}_{i}^{a}$ is to the task requirements $\mathbf{x}_{k}^{a}$ in $\mathcal{M}_{a}$.

Assuming independence between available tasks, the complementarity between agents $i$ and $j$ is the probability $p_{ij}$ they can co-execute at least one task:
\begin{equation}
\label{eq:compl11}
p_{ij} = 1 - \prod_{k} \left(1- p(i,j|k)\right).
\end{equation}

Equations~(\ref{eq:compl1}) and (\ref{eq:compl11}) serve as a foundation for the general framework for modeling and learning complementarity representations in real networks, which we summarize in Fig.~\ref{fig:compl}{\bf a}. From the modeling perspective, the complementarity framework can be used to generate both bipartite (when both agents and tasks are present) and conventional networks (when only agents are considered). To generate a complementarity-based model network one needs to define latent spaces $\mathcal{M}_{1}$ and $\mathcal{M}_{2}$, connection probabilities $r_{1}(d)$ and $r_{2}(d)$, and the mechanism of distributing nodes in tasks in the latent space. Since connections between the nodes are functions of the node positions in the latent spaces, and network links are established independently of one another, the complementarity framework belongs to the class of network models with hidden variables~\cite{boguna2003class,kitsak2011hidden} -- with node positions serving the roles of hidden variables -- allowing for analytical treatment.

From the learning perspective, obtaining a complementarity representation of a real network includes learning latent spaces $\mathcal{M}_{1}$ and $\mathcal{M}_{2}$, connection probability functions $r_{1}(d)$ and $r_{2}(d)$, as well as the coordinates of agents and tasks in the two latent spaces from the observed adjacency matrix $A$ of the network of interest.

\section{Minimal Complementarity Framework}
\label{sec:minimal_compl}

From practical considerations, learning the principled complementarity representation may prove suboptimal. One reason is that tasks are either poorly defined in many systems, for instance, in the case of molecular interaction networks, or not observable, like in the case of semantic networks.  Another reason is that the original framework contains too many ``degrees of freedom", which may lead to overfitting when learning representations of real systems. Therefore, our next step is to reduce the principled framework that includes both agents and tasks to a more practical one that only includes agents.

If connection probabilities are small, $p(i,j|k) \ll 1$, the largest contribution to $p_{ij}$ in Eq.~(\ref{eq:compl11}) comes from the linear terms in $p(i,j|k)$, resulting in
\begin{equation}
\label{eq:compl12}
p_{ij} \approx \sum_{k} p(i,j|k).
\end{equation}
In the case the number of tasks $T$ is large, we can replace individual tasks with a certain mean field, treating them as drawn independently at random from a joint pdf $\rho_{T}  \left(\mathbf{x}^{1}, \mathbf{x}^{2} \right)$. In this case, we can replace the sums over individual tasks in Eq.~(\ref{eq:compl12}) by the integrals over the volumes of $\mathcal{M}_{1}$ and $\mathcal{M}_{2}$, obtaining, in the leading order,
\begin{equation}
\label{eq:compl22}
p_{ij} = p_{ij}\left(\mathbf{x}_{i}^{1}, \mathbf{x}_{i}^{2}; \mathbf{x}_{j}^{1},  \mathbf{x}_{j}^{2}\right)= \int_{\mathcal{M}_1}\int_{\mathcal{M}_2} {\rm d} \mathbf{x}^{1}{\rm d}  \mathbf{x}^{2} \rho_{T}\left( \mathbf{x}^{1}, \mathbf{x}^{2}\right)  \left[r_{1}\left(d\left[\mathbf{x}_{i}^{1}, \mathbf{x}^{1}\right]\right)r_{2}\left(d\left[\mathbf{x}_{j}^{2}, \mathbf{x}^{2}\right]\right)  +  i \leftrightarrow j \right],
\end{equation}
where the second term is obtained from the first one by swapping indices $i$ and $j$.

The second simplification can be made by assuming a particular functional form for $\rho_{T}  \left(\mathbf{x}^{1}, \mathbf{x}^{2} \right)$. To this end, the simplest case is that of uncorrelated task coordinates, $\rho_{T}\left(\mathbf{x}^{1}, \mathbf{x}^{2}\right) = \rho_{1}\left(\mathbf{x}^{1}\right)\rho_2\left(\mathbf{x}^{2}\right)$. It is straightforward to verify that this choice leads to a trivial degenerate model, where the collaboration probability of any two agents $i$ and $j$ is proportional, in the leading order, to the product of their individual productivities. 

A more interesting case, as empirical studies seem to suggest~\cite{Sinatra2015century,Pan2012evolution,Gates2019nature}, is where there is a strong correlation between the task coordinates in $\mathcal{M}_{1}$ and $\mathcal{M}_{2}$. In this case, the simplest choice for $\rho_{T}\left(\mathbf{x}^{1},\mathbf{x}^{2}\right)$ is obtained by a deterministic injective continuous relationship $f:\mathcal{M}_1 \to \mathcal{M}_2$ between skills in the two spaces: $\mathbf{x}^{2} = f (\mathbf{x}^{1})$. 
In this case, task coordinates  $\mathbf{x}_{k}^{1}$ in $\mathcal{M}_{1}$ uniquely determines the task coordinates $\mathbf{x}_{k}^{2}$ in $\mathcal{M}_{2}$, and  
the pdf of the task distribution is

\begin{equation}
\rho_T(\mathbf{x}^1, \mathbf{x}^{2}) = \rho(\mathbf{x}^{1}) \delta \left(\mathbf{x}^{2} - f \left(\mathbf{x}^{1}\right)\right),
\label{eq:rho_corr}
\end{equation}
where $\delta(\mathbf{x})$ is a multi-dimensional Dirac delta function. 
  
The strong correlation between task coordinates provides an effective mapping between skills in $\mathcal{M}_1$ and $\mathcal{M}_2$.
Indeed, for every point $\mathbf{y}$ in $\mathcal{M}_1$, there is a unique point $\mathbf{x} = f \left(\mathbf{y}\right)$ in $\mathcal{M}_2$ and vice-versa. Using this property and the explicit form of $\rho_T$ given by Eq.~(\ref{eq:rho_corr}), we integrate out  $\mathbf{x}^{2}$, obtaining the collaboration probability $p_{ij}$ as a function of type 1 skills in $\mathcal{M}_{1}$, $\mathbf{x}_{i}^{1}$ and  $\mathbf{x}_{j}^{1}$, and the images of type 2 skills in  $\mathcal{M}_{1}$, $\mathbf{y}_{i}^{1} \equiv f^{-1}\left(\mathbf{x}_{i}^{2} \right)$ and  $\mathbf{y}_{j}^{1}  \equiv f^{-1}\left(\mathbf{x}_{j}^{2} \right)$:
\begin{equation}
\label{eq:compl2}
p_{ij}\left(\mathbf{x}_{i}^{1}, \mathbf{y}_{i}^{1}; \mathbf{x}_{j}^{1}, \mathbf{y}_{j}^{1}\right)= \int_{\mathcal{M}_1}{\rm d} \mathbf{x}^{1} \rho\left(\mathbf{x}^1\right) \left[r_{1}\left(d\left[\mathbf{x}_{i}^{1},\mathbf{x}^{1}\right]\right)r_{2}\left(d\left[f\left(\mathbf{y}_{j}^{1}\right),f\left(\mathbf{x}^{1}\right)\right]\right)  +  i \leftrightarrow j \right],
\end{equation}
see Fig.~\ref{fig:compl}{\bf b}. Finally, if tasks are distributed uniformly in $\mathcal{M}_{1}$, we get
\begin{equation}
\label{eq:compl21}
p_{ij}\left(\mathbf{x}_{i}^{1}, \mathbf{y}_{i}^{1}; \mathbf{x}_{j}^{1}, \mathbf{y}_{j}^{1}\right)= \frac{1}{{\rm vol}_{\mathcal{M}_{1}}} \int_{\mathcal{M}_1}{\rm d} \mathbf{x}^{1}  \left[r_{1}\left(d\left[\mathbf{x}_{i}^{1},\mathbf{x}^{1}\right]\right)r_{2}\left(d\left[f\left(\mathbf{y}_{j}^{1}\right),f\left(\mathbf{x}^{1}\right)\right]\right)  +  i \leftrightarrow j \right],
\end{equation}
where ${\rm vol}_{\mathcal{M}_{1}}$ is the volume of $\mathcal{M}_{1}$.

At this point, we can make another important observation for Eq.~(\ref{eq:compl21}). The collaboration between agents $i$ and $j$ is likely if either $\mathbf{x}_{i}^{1}$ is close to $\mathbf{y}_{j}^{1}$, or  $\mathbf{x}_{j}^{1}$ is close to $\mathbf{y}_{i}^{1}$, or both, see Fig.~\ref{fig:compl}{\bf b}. This is the case since $r_{1}(d)$ and $r_{2}(d)$ are both decreasing functions of $d$. Then, the largest contributions to the first term in Eq.~(\ref{eq:compl21}) are those when both $d\left[\mathbf{x}_{i}^{1},\mathbf{x}^{1}\right]$ and $d\left[f\left(\mathbf{y}_{j}^{1}\right),f\left(\mathbf{x}^{1}\right)\right]$ are small at the same time. This is the case when $\mathbf{x}_{i}^{1} = \mathbf{y}_{j}^{1}$. By the same argument, the contributions to the second integral in Eq.~(\ref{eq:compl21}) are maximized when $\mathbf{x}_{j}^{1} = \mathbf{y}_{i}^{1}$.

In summary, we have made two observations. The first one is that the strong correlation between task skills in $\mathcal{M}_1$ and $\mathcal{M}_{2}$ provides an effective mapping between the two spaces, allowing us to project type 2 skills from  $\mathcal{M}_{2}$ onto $\mathcal{M}_{1}$ and vice versa. In this case, one can reduce the complementarity framework to one latent space. If this space is $\mathcal{M}_{1}$, type 2 skills in it are the images of type 2 skills in $\mathcal{M}_{2}$, $\mathbf{y}^{1} = f^{-1}\left(\mathbf{x}^{2}\right)$. The second conclusion is that the complementarity between two agents $i$ and $j$, quantified by their collaboration probability, is maximized when either the type 1 skill of $i$ ($\mathbf{x}_{i}^{1}$) is close to the type 2 skill of $j$ ($\mathbf{y}_{j}^{1}$), or the type 2 skill of $i$ ($\mathbf{y}_{i}^{1}$) is close to the type 1 skill of $j$  ($\mathbf{x}_{j}^{1}$), or when both are close.

These findings allow us to guess a simpler form of the collaboration probability. Without loss of generality, we consider the projection of the generalized complementarity framework onto the latent space $\mathcal{M}_{1}$. Since there is only one latent space after the projection, we can simplify the notation by dropping upper indices for node coordinates. Each agent $i$ is characterized by two features or skills in latent space $\mathcal{M}$, which we refer to as $\mathbf{x}_{i}$, $\mathbf{y}_{i}$. Consistent with the first observation that strong correlations between task features provide an effective mapping between the two manifolds, we postulate that the features of both types are represented by points in the same metric space $\mathcal{M}$: the features $\mathbf{x}_{i}$ are native in the space, while the features $\mathbf{y}_{i}$ are images projected by the tasks from the second space.

Satisfying the second observation, we postulate that the complementarity between any two agents $i$ and $j$ is a function of the distances between their skills of different types, $d \left[\mathbf{x}_{i}, \mathbf{y}_{j} \right]$ and $d \left[\mathbf{y}_{i}, \mathbf{x}_{j} \right]$,  in $\mathcal{M}$:
\begin{equation}
p\left(\mathbf{x}_{i}, \mathbf{y}_{i}; \mathbf{x}_{j}, \mathbf{y}_{j}\right)=  1 - \left[1-p\left(d \left[\mathbf{x}_{i}, \mathbf{y}_{j} \right]\right)\right] \left[1 - 
p\left(d \left[\mathbf{y}_{i}, \mathbf{x}_{j} \right]\right)\right],
\label{eq:compl3}
\end{equation}
where $p:R^{+} \to \left[0,1\right]$ is a decreasing function.
Equation~(\ref{eq:compl3}) prescribes that the complementarity $p_{ij}\left(\mathbf{x}_{i}, \mathbf{y}_{i}; \mathbf{x}_{j}, \mathbf{y}_{j}\right)$ is maximized when $\mathbf{x}_{i}$ is close to $\mathbf{y}_{j}$ or $\mathbf{y}_{i}$ is close to $\mathbf{x}_{j}$, or both. Complementarity $p_{ij}$ in (\ref{eq:compl3}) can be interpreted as the union of two probabilities: the complementarity between two agents is possible due to either the complementarity of skills $\mathbf{x}_{i}$ and $ \mathbf{y}_{j}$, or $\mathbf{y}_{i}$ and $\mathbf{x}_{j}$, or both. 

A very important observation that we can make based on Eq.~(\ref{eq:compl3}) is that the complementarity can also model similarity in the special case where the two points of a given node are the same, $\mathbf{x}_{i} = \mathbf{y}_{i}$. If this is the case for a single node $i$, we can think of this node as self-complementary. In the context of molecular interactions, we are dealing with a molecule capable of self-interacting. In the context of interdisciplinary collaborations, a self-complementary agent possesses sufficient skills in the two domains to solve tasks on their own. At the same time, self-complementary nodes can interact with other nodes that are complementary to them. If all network nodes are self-complementary, the complementarity framework reduces to a similarity framework. In this case, every node is effectively characterized by a single point in $\mathcal{M}$, since $\mathbf{x}_{i} = \mathbf{y}_{i}$ for $i = 1,\ldots,N$, and the complementarity between any two nodes $i$ and $j$ in Eq.~(\ref{eq:compl3}) reduces to a function of the distance between the corresponding points, 
$p_{ij} = p\left(\mathbf{x}_{i}; \mathbf{x}_{j}\right)=2 p\left(d \left[\mathbf{x}_{i}, \mathbf{x}_{j} \right]\right) - p\left(d \left[\mathbf{x}_{i}, \mathbf{x}_{j} \right]\right)^{2}$.

\begin{figure}
\includegraphics[width=5in]{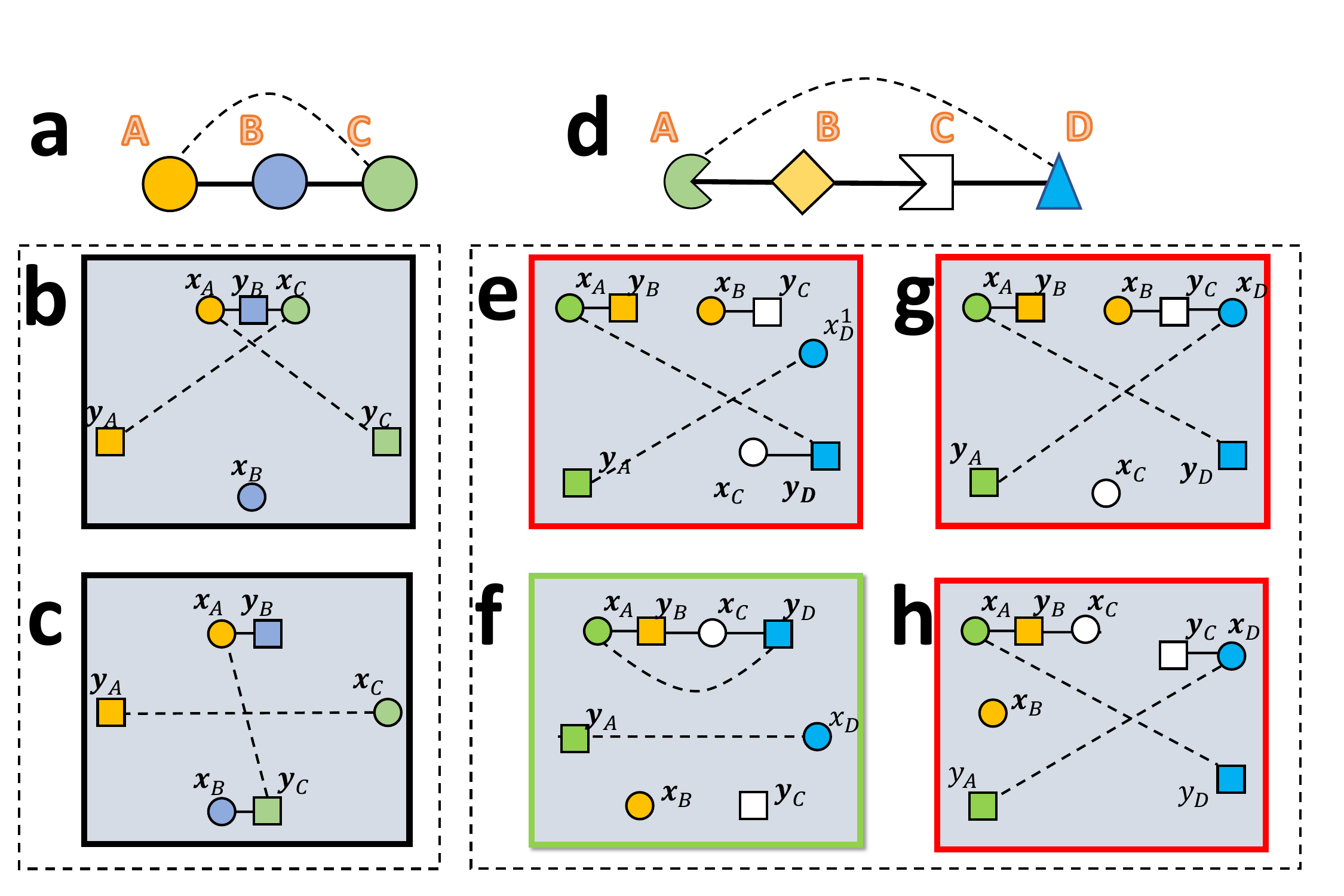}
\caption{\footnotesize (color online) {\bf Minimal complementarity framework does not support the triangle closure rule and may support the diamond-closure rule.} {\bf a}, The triangle closure rule suggests that node $A$ is connected to node $C$ in the toy network, forming the triangle subgraph. There are four complementarity representations that may lead to the wedge toy network in $\bf a$. Two representations are shown in ${\bf b}$ and ${\bf c}$, the other two are obtained from ${\bf b, c}$ by swapping the features of each node, $\mathbf{x}_{i}  \leftrightarrow \mathbf{y}_{i}$. Connections between node pairs $A$-$B$ and $B$-$C$ are possible  due to ({\bf b}) latent-geometric proximity of points $\mathbf{x}_{A}$, $\mathbf{y}_{B}$, and $\mathbf{x}_{C}$ or ({\bf c}) latent geometric proximity of point pairs $\mathbf{x}_{A}$ and $\mathbf{y}_{B}$, and $\mathbf{x}_{B}$ and $\mathbf{y}_{C}$. Neither configuration imposes constraints on distances between point pairs $\mathbf{x}_{A}$ $\mathbf{y}_{C}$, and $\mathbf{y}_{A}$ and $\mathbf{x}_{C}$, which are relevant to the formation of the $A$-$C$ tie. {\bf d} The diamond closure rule suggest that node $A$ is connected to $D$ completing the diamond in the $A$-$B$-$C$-$D$ toy network. There are 8 latent-geometric configurations corresponding to the toy network in {\bf d}. Four configurations are shown in {\bf e-h}, the remaining four are obtained from configurations in {\bf e-h} by swapping the features of each node, $\mathbf{x}_{i}  \leftrightarrow \mathbf{y}_{i}$. Note that only two out of possible eight configurations, {\bf f}  and its counterpart, imply the $A$-$D$ connection.}
\label{fig:2}
\end{figure}

We note that the proposed minimal complementarity framework is consistent with the non-transitivity of complementarity: node $A$ being complementary to $B$ and $B$ complementary to $C$ do not imply that $A$ is complementary to $C$. Indeed, within the minimal complementarity framework the formation of the $A$-$B$-$C$ toy wedge network, Fig.~\ref{fig:2}{\bf a}, is possible in two cases: (i) points $\mathbf{x}_{A}$, $\mathbf{y}_{B}$, and $\mathbf{x}_{C}$ are close to each other, Fig.~\ref{fig:2}{\bf b}, or (ii) point $\mathbf{x}_{A}$ is close to point $\mathbf{y}_{B}$, and $\mathbf{x}_{B}$ is close to $\mathbf{y}_{C}$, Fig.~\ref{fig:2}{\bf c}. Neither case creates constraints on distances $d\left(\mathbf{x}_{A}, \mathbf{y}_{C}\right)$ and $d\left(\mathbf{y}_{A}, \mathbf{x}_{C}\right)$, relevant for the formation of the $A$-$C$ link.

This observation is consistent with a recent result in the prediction of protein-protein interactions, Ref.~\cite{kovacs2019network}, where the authors argued that the complementarity principles suppress even-length network-based paths between interacting proteins but at the same time promote the appearance of odd-length network-based paths. To complete the comparison, we asked if network-based paths of length $\ell=3$ in the minimal complementarity framework imply a connection between the path endpoints. To answer this question, we considered the most likely geometric configurations leading to an $\ell=3$ path, Fig.~\ref{fig:2}{\bf d}. As seen in Figs.~\ref{fig:2}{\bf e,f,g,h}, two out of eight possible configurations imply the appearance of a link between the endpoints of the $\ell=3$ path. In other words, while the minimal complementarity framework does not suppress connections between the endpoints of $\ell=3$ paths, it also does not strongly impose them. We revisit this question in Section~\ref{sec:modeling}, where we assess the densities of 3- and 4-cycles in a synthetic network generated with the minimal complementarity framework.

In our analysis, we use point representations for similarity-based fields: agents are defined as points, and similarities are quantified as distances between the points: the shorter the distance, the higher the similarity. While the distance-based definition of similarity is popular among social sciences, physics, and network science communities, there exists an alternative vector representation of similarity, which is more common in computer sciences. In a vector representation, each network node $i$ is represented as a vector $\mathbf{v}_{i}$ and the similarity between any two nodes $i$ and $j$ is quantified by the inner product of the two vectors $\mathbf{v}_{i} \cdot \mathbf{v}_{j}$. We chose distance-based representations because they are identical for all metric spaces $\mathcal{M}$. While vector representations are simple in vector spaces, e.g., {\it Euclidean} spaces, extending these representations to non-vector spaces, e.g., hyperbolic spaces, is not easy.

It should be straightforward, starting with the same first principles, to work out the complementarity framework using vector representations. Here we conjecture that, similar to the point representation, vector-based complementarity frameworks will characterize each node $i$ by at least two vectors, which we call $\mathbf{u}_{i}$ and $\mathbf{v}_{i}$, such that the complementarity of two nodes $i$ and $j$ is a function of the inner products of vectors of different types.
\begin{equation}
p_{ij} = \rho \left(\mathbf{u}_{i} \cdot \mathbf{v}_{j}; \mathbf{v}_{i} \cdot \mathbf{u}_{j}\right)
\label{eq:vector_compl}
\end{equation}

By examining Eq.~(\ref{eq:vector_compl}), we realize that some network embedding methods previously developed in machine learning and natural language processing (NLP) communities do represent network nodes or vectors by two vectors each. The most basic is, arguably, the singular value decomposition (SVD), which can be used to factorize the network adjacency matrix $A$ as $A = U \Sigma V^{T}$, where $\Sigma$ is the diagonal matrix with the singular values of $A$ on the diagonal, and $U$ and $V$ can be regarded as two vector representations for each network node. The Global Vectors for Word Representation (GloVe) unsupervised NLP approach represents each word $i$ by two vectors $\mathbf{u}_{i}$ and $\mathbf{v}_{i}$, referred to as the target and context vectors, respectively~\cite{pennington2014glove}. GloVe quantifies the frequency of finding word $i$ in the context of word $j$ by a function of the inner product of the corresponding vectors, $\mathbf{u}_{i} \cdot \mathbf{v}_{j}$. Recently, the GloVe method has been adapted for network embedding, known as Global Vectors for Node Representations (GVNR)~\cite{brochier2019global}. Since SVD and GVNR use two-vector-per-node representations, they have the potential to satisfactorily learn complementarity representations. Therefore, we analyze the complementarity framework alongside the SVD and GVNR methods.

In summary, the minimal complementarity framework can be used in both modeling synthetic complementarity networks and learning complementarity representations of real networks. We discuss them in the following sections. Both complementarity representations can be viewed as a certain juxtaposition of two or more similarity representations. Indeed, we defined complementarity as the ability of agents to co-execute interdisciplinary tasks in two distinct similarity-based spaces. It is the manifestation of tasks in the corresponding similarity-based fields that connects the agents and defines the nature of complementarity.

\section{Synthetic complementarity-based models.}
\label{sec:modeling}
The minimal complementarity framework can be realized for any latent space $\mathcal{M}$ and any decreasing connection probability function $p(d)$. We anticipate that for the best results both $\mathcal{M}$ and $p(d)$ need to be tailored to or learned from the network $G$ of interest. For brevity, we leave the task of determining $\mathcal{M}$ and $p(d)$ for future work and focus on the example of using the hyperbolic disk, $\mathcal{M} = \mathbb{H}^{2}$, and the case of $p(d)$ given by the sigmoid shape. We refer to the resulting model as the Complementarity Random Hyperbolic Graph (CRHG). 

The CRHG is inspired by the success of the similarity-based random hyperbolic graph (RHG)~\cite{Krioukov2010hyperbolic}, which uses a representation of one point per node, in modeling properties of real similarity-based networks. It is now well-understood that RHGs reproduce many topological properties of real networks, such as sparsity, scale-free degree distribution, strong clustering coefficient, hierarchical structure, and self-similarity~\cite{Serrano2008,Krioukov2010hyperbolic,Papadopoulos2012popularity,zuev2015emergence}. In the RHG, nodes are sprinkled as points into a hyperbolic disk $\mathbb{H}^{2}$ and pairwise connections are established independently with probabilities that are decreasing functions of the distances between the corresponding points in $\mathbb{H}^{2}$.

Following the minimal complementarity framework, we demand that every node of the CRHG is represented by two points in $\mathbb{H}^{2}$ and connections between nodes are established independently according to Eq.~(\ref{eq:compl3}). In more precise terms, each node $i$ is characterized by two points $\mathbf{x}_{i} \equiv \{ r^{1}_{i}, \theta^{1}_{i}\}$ and
 $\mathbf{y}_{i} \equiv \{ r^{2}_{i}, \theta^{2}_{i}\}$  in  $\mathbb{H}^{2}$ and the connection probability is
\begin{equation}
\label{eq:compl4}
p_{ij}= p\left(d \left[\mathbf{x}_{i}, \mathbf{y}_{j} \right]\right) + p\left(d \left[\mathbf{x}_{j}, \mathbf{y}_{i} \right]\right)  - p\left(d \left[\mathbf{x}_{i}, \mathbf{y}_{j} \right]\right) \times p\left(d \left[\mathbf{x}_{j}, \mathbf{y}_{i} \right]\right) ,
\end{equation}
where $d \left[\mathbf{x}_{i}, \mathbf{y}_{j} \right]$ is the distance between points $\mathbf{x}_{i} $ and $\mathbf{y}_{j} $ in  $\mathbb{H}^{2}$: 

\begin{eqnarray}
\cosh d \left[r^{1}_{i}, \theta^{1}_{i}; r^{2}_{j}, \theta^{2}_{j}\right] &= &\cosh r_{i}^{1} \cosh r_{j}^{2} - \sinh r_{i}^{1} \sinh r_{j}^{2} \cos \Delta \theta^{12}, \\
 \Delta \theta^{12} &\equiv& \pi - \|\pi - \|\theta_{i}^{1} - \theta_{j}^{2} \| \|.
\label{eq:hypercos}
\end{eqnarray}

We can distinguish two special cases of the CRHG. In the case where the two points $\mathbf{x}_{i} $ and $\mathbf{y}_{j}$ are selected independently for each node $i$ we have the \emph{pure} complementarity model. On the other hand, if the two points per node are equal to each other,  $\mathbf{x}_{i} = \mathbf{y}_{j}$, the CRHG reduces to a similarity model that is very close to the classical RHG, with the only difference that each connection is established with two independent attempts. In a more general setting, a certain mixture CRHG model is obtained if every node $i = 1,\ldots,N$ with probability $f \in (0,1)$ has two independent points $\{ r^{1}_{i}, \theta^{1}_{i}\}$ and $\{ r^{2}_{i}, \theta^{2}_{i}\}$ and with probability $1-f$ has identical points $r^{1}_{i} = r^{2}_{i}$ and $\theta^{1}_{i} = \theta^{2}_{i}$.

Below we consider the pure CRHG, $f=0$, and postpone the analysis of the general $f \neq 0$ case for future work. Inspired by the RHG, we assume that the connection probabilities in Eq.~(\ref{eq:compl4}) are of the sigmoid shape:
\begin{equation}
p(d) = \frac{1}{1 + e^\frac{d-R}{2T}},
\label{eq:fd}
\end{equation}
where $T \in (0,1)$ is the temperature parameter controlling the relevance of long-distance connections, and $R > 0$ is the radius of the hyperbolic disk. Small values of $T$ result in models where most connections are established at distances of at most $R$. As $T$ increases, connections are more likely to take place at distances $d > R$.

The node coordinates are drawn uniformly at random in $\mathbb{H}^{2}$ as
 \begin{eqnarray}
 \theta^{1,2}_{i} & \leftarrow& U(0, 2\pi), \\
 r^{1,2}_{i} &\leftarrow &\frac{\sinh \left( \alpha r^{t} \right) }{ \cosh (\alpha R) - 1},~r_{i}^{1,2}\in[0,R]
 \end{eqnarray}
for $i = 1,\ldots, N$. Here $R > 0$ is the radius of the hyperbolic disk and $\alpha > 1$ is a parameter controlling the density of nodes.

Using the hidden variable formalism~\cite{boguna2003class}, we find that the degree distribution of the CRHG follows a power law, $P(k) \sim k ^{-\gamma}$
with $\gamma = 2\alpha + 1$. Indeed, the expected degree of a node characterized by the coordinates $\{r^{1}_{i}, \theta^{1}_{i}, r^{2}_{i}, \theta^{2}_{i}\}$, to the leading order, is given by
\begin{equation}
\overline{k}\left(r^{1}_{i}, \theta^{1}_{i}, r^{2}_{i}, \theta^{2}_{i}\right) =\overline{k}^{\mathbb{H}}(r^{1}_{i}) + \overline{k}^{\mathbb{H}}(r^{2}_{i}),
\end{equation}
where $\overline{k}^{\mathbb{H}}(r)$ is the expected degree of a node with radial coordinate $r$ in the RHG:
\begin{equation}
\overline{k}^{\mathbb{H}}(r) = \frac{4N\alpha}{2\alpha - 1} \frac{T}{\sin \pi T}  e^{-r/2},
\end{equation}
see Ref.~\cite{Krioukov2010hyperbolic}. Then, the degree distribution of the CRHG is nothing else but the convolution of the degree distributions of two RHGs: 
\begin{equation}
  P(k) = \sum_{k' = 0}^{k} P^{\mathbb{H}}(k-k') P^{\mathbb{H}}(k') \sim k^{-\gamma},
 \label{eq:chrg_pk}
\end{equation}
where $\gamma = 2\alpha + 1$.

Similarly, the expected number of common neighbors between nodes $i$ and $j$ can be expressed, to the leading order, as
\begin{equation}
\overline{m}_{ij} = \overline{m}^{\mathbb{H}}\left(r^{1}_{i},\theta^{1}_{i};r^{1}_{j},\theta^{1}_{j}\right) + \overline{m}^{\mathbb{H}}\left(r^{2}_{i},\theta^{2}_{i};r^{2}_{j},\theta^{2}_{j}\right),
\end{equation}
where $\overline{m}^{\mathbb{H}}\left(r_{i},\theta_{i};r_{j},\theta_{j}\right)$ is the expected number of common neighbors of nodes with hidden variables $\{r_i,\theta_i\}$ and $\{r_j,\theta_j\}$ in an RHG. Then, the probability $P(m)$ of two randomly chosen nodes in the CRHG to have exactly $m$ common neighbors is the convolution of those in the RHG model
\begin{equation}
  P(m) = \sum_{m' = 0}^{m} P^{\mathbb{H}}(m-m') P^{\mathbb{H}}(m') \sim m^{-\tau},
   \label{eq:chrg_pm}
\end{equation}
since $P^{\mathbb{H}}(m) \sim m ^{-\tau}$, where exponent $\tau = \tau(\gamma, T) >2$, as documented both in empirical~\cite{iamnitchi2004small,burgos2008two} and theoretical~\cite{kitsak2017latent} studies.

\begin{figure}
\includegraphics[width=2.5in]{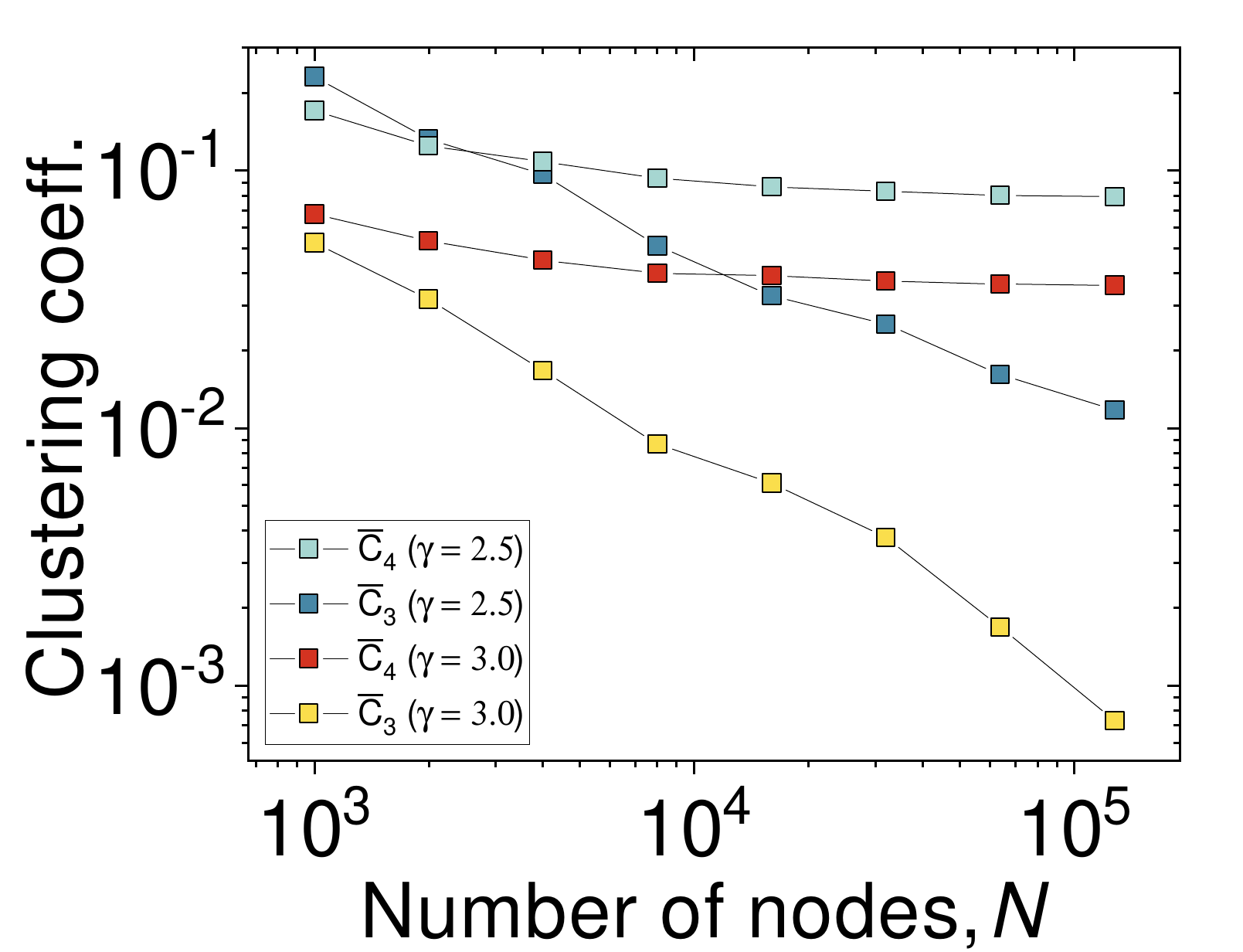}
\caption{\footnotesize (color online) {\bf CRHG models are characterized by vanishing densities of $3$-cycles, and non-vanishing densities of $4$-cycles.} Average degree-dependent clustering coefficients $\bar{C}_{3}$ and $\bar{C}_{4}$ as a function of network size $N$. Networks are CRHG models with parameters $\alpha = 0.75$ ($\gamma = 2.5$) and $\alpha = 1.0$ ($\gamma = 3.0$). In all CRHG networks, we set  $\langle k \rangle = 20$, and $T=0.5$.  }
\label{fig:c3_c4}
\end{figure}

When formulating the complementarity framework in Section~\ref{sec:minimal_compl}, we conjectured that it should suppress the appearance of odd-length cycles and, conversely, promote the appearance of even-length cycles.  Consistent with our expectations, we observe that the average clustering coefficient vanishes as a function of size $N$, while bipartite generalizations of the clustering coefficient are independent of $N$ in the CRHG model, Fig.~\ref{fig:c3_c4}.

\begin{figure}
\includegraphics[width=6in]{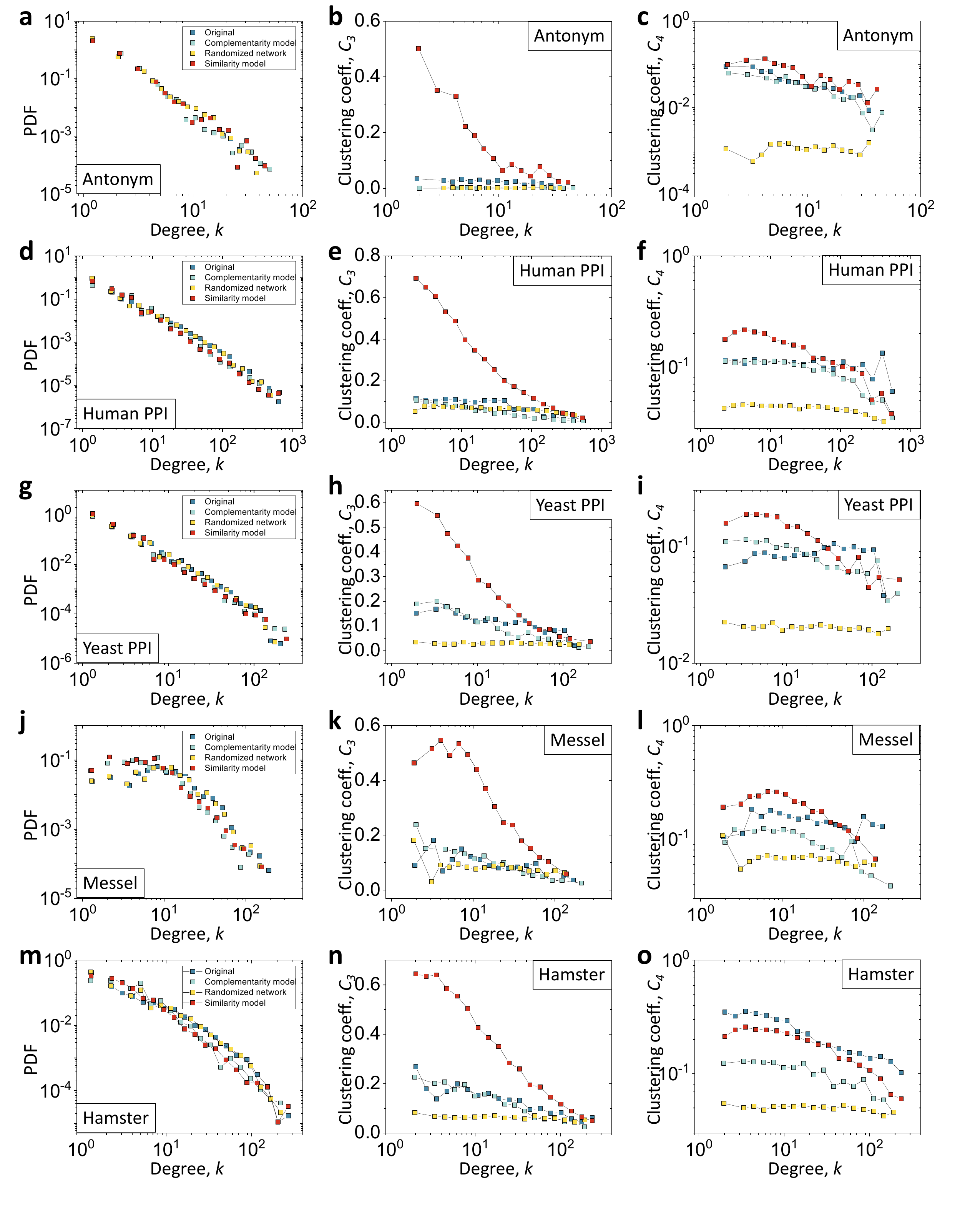}
\caption{\footnotesize (color online) {\bf Topological properties of complementarity-driven networks.} Shown are degree distributions $P(k)$, degree-based clustering $C_{3}$ and bipartite clustering $C_{4}$ coefficients for {\bf a-c} antonym, {\bf d-f} human PPI, {\bf g-i} yeast PPI, {\bf j-l} Messel, and {\bf m-o} Hamster networks. Topological properties of real networks (blue) are compared with those of their degree-preserved randomizations (yellow), CRHGs (light blue), and RHGs (red).} 
\label{fig:topology_comparison}
\end{figure}

To assess how well the CRHG can reproduce topological properties of real networks, we considered the networks of antonyms, human~\cite{Das2012,luck2020reference} and yeast~\cite{Yu2008} protein-protein interaction (PPI) networks, the Messel shell food web~\cite{dunne2014highly}, and the social network of the website hamsterster.com~\cite{kunegis2013konect}, see Table~\ref{app:table} and Appendix~\ref{sec:net_properties}. We measured degree distribution $P(k)$, degree-based clustering coefficient $\overline{C}_{3}(k)$ and its bipartite counterpart $\overline{C}_{4}(k)$~\cite{kitsak2017latent} for these networks. We compared the properties of the real networks to their degree-preserving randomizations, as well as synthetic graphs from the CRHG and RHG models, Fig.~\ref{fig:topology_comparison}. We observe that both the CRHG (complementarity) and RHG (similarity) can adequately model the degree distributions of real networks, Fig.~\ref{fig:topology_comparison}{\bf a, d, g, j, m}. All studied real networks are characterized by small values of the clustering coefficient that are comparable to those of their degree-preserving randomizations and corresponding CRHG models, Fig.~\ref{fig:topology_comparison}{\bf b, e, h, k, n}. Similarity-based RHGs, on the other hand, tend to over-inflate with triangles, resulting in substantially higher clustering coefficients. Both the RHG and CRHG models are capable of generating networks with significant densities of 4-cycles, as measured by $\overline{C}_{4}(k)$, Fig.~\ref{fig:topology_comparison}{\bf c, f, i, l, o}.

\section{Learning complementarity representations of real networks.}

After establishing that the CRHG model can generate synthetic networks with topological properties similar to real complementarity-based systems, we next use CRHG to learn hyperbolic complementarity representations of these real systems. 

In a hyperbolic complementarity representation, every node $i = 1,\ldots,N$ in a network $G$ of interest is characterized by two points $\mathbf{x}_{i} \equiv 
\{ r^{1}_{i}, \theta^{1}_{i}\}$ and $\mathbf{y}_{i} \equiv \{ r^{2}_{i}, \theta^{2}_{i}\}$ in hyperbolic disk $\mathbb{H}^{2}$. To learn complementarity representations, we use the maximum likelihood estimation (MLE) approach. Given a network $G$ with adjacency matrix $A$, we aim to find $\{\mathbf{x}_{i}\}$ and $\{\mathbf{y}_{i}\}$ for $i = 1,\ldots,N$ maximizing the likelihood $\mathcal{L}$ that $G$ is generated from the CRHG model with the given node assignment. Due to the independence of links in the CRHG, the thought likelihood is given by

\begin{equation}
\mathcal{L}\left(  a_{ij}| \{r^{1}_{i}, \theta^{1}_{i}; r^{2}_{i}, \theta^{2}_{i} \}, \mathcal{P}\right) = \prod_{i < j} \left[ q p_{ij}\right]^{a_{ij}}  \left[1 -q p_{ij}\right]^{1 - a_{ij}},
\label{eq:likelihood}
\end{equation}
where $a_{ij}$ are the elements of the network adjacency matrix $A$, connection probabilities $p_{ij}$  are prescribed by Eq.~(\ref{eq:compl4}), and $q\in[0,1]$ is the presence probability of each link. We use $q=1$ when learning representations of fully-observable networks. If, on the other had, a fraction $\lambda$ of network links are unobserved, we set $q=1-\lambda$. Our complementarity learning algorithm, which we call Complementarity Hyper Link (CHL) embedder is identical to the HyperLink embedder~\cite{kitsak2020link}, modulo the likelihood function $\mathcal{L}$ of Eq.~(\ref{eq:likelihood}), see Section~\ref{app:embedding}.

Using the CHL embedder, we learned complementarity representations of the five complementarity-based systems. To assess the quality of  the CHL and the accuracies of the obtained complementarity representations, we used them to predict missing network links. To do so, we removed a fraction of $1-q = 0.5$ links in each network uniformly at random and then learned complementarity representations of the resulting truncated networks. We then ranked all missing link candidates (unconnected node pairs) by the the sum of the complementarity distances between the corresponding points:
\begin{equation}
{\rm rank}_{ij} = d\left(\mathbf{x}^{1}_i, \mathbf{x}^{2}_j\right) + d\left(\mathbf{x}^{2}_i, \mathbf{x}^{1}_j\right),
\end{equation}
such that the smaller the rank is, the higher is the chance of a missing link.

\begin{figure}
\includegraphics[width=6in]{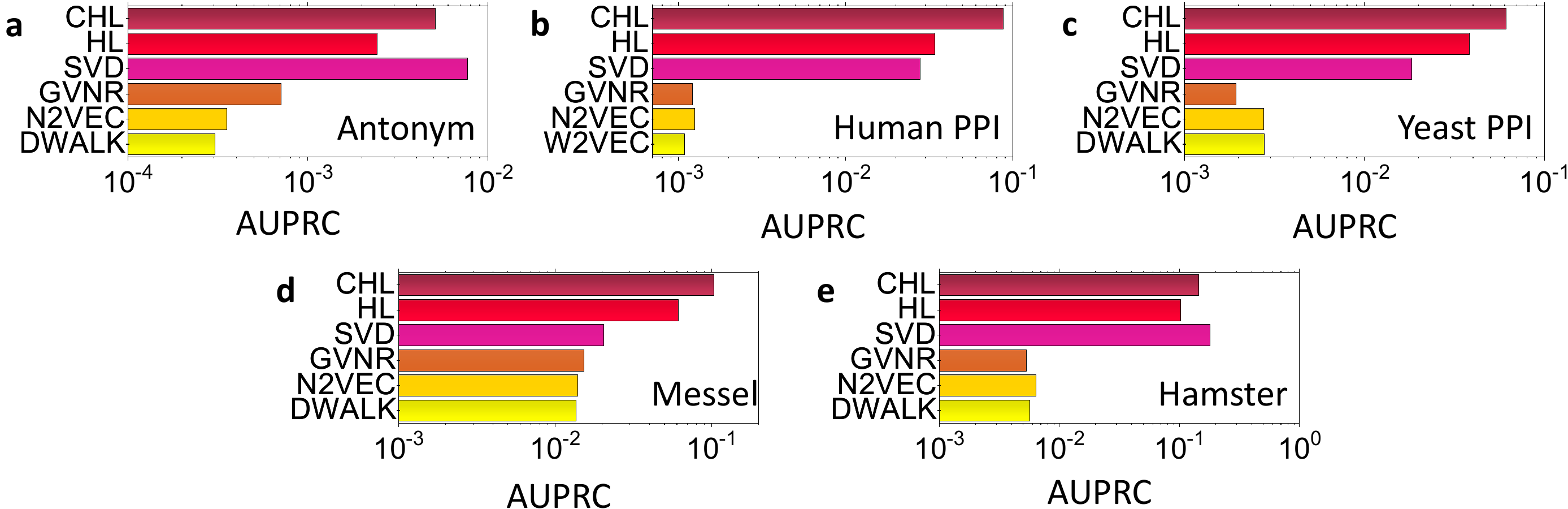}
\caption{\footnotesize (color online) {\bf Embedding quality as quantified by the link prediction accuracy.} Link prediction results obtained with the Complementarity HyperLink (CHL) are compared to ordinary HyperLink (HL), singular value decomposition (SVD), node2vec (N2VEC), and deepwalk (DWALK) in {\bf a} the network of antonyms (Antonyms), {\bf b} human protein-protein interaction network (Human PPI), {\bf c} \emph{S. cerevisiae}  protein-protein interaction network (Yeast PPI), {\bf d} Messel food web (Messel), and ({\bf d}) Hamsterster social network (Hamster). All experiments correspond to a fraction $1-q=0.5$ of removed links, the accuracy of link prediction is evaluated using the area under the Precision-Recall characteristic (AUPRC). In all embedding {\it Euclidean}-basedmethods  we set the latent space dimensionality to $D=256$. We used $100$ epochs in the N2VEC and DWALK embeddings and $10$ epochs in the GVNR embeddings. For link prediction with SVD we truncated the obtained decomposition vectors at $D=256$.
\label{fig:31}}
\end{figure}

\begin{figure}
\includegraphics[width=6in]{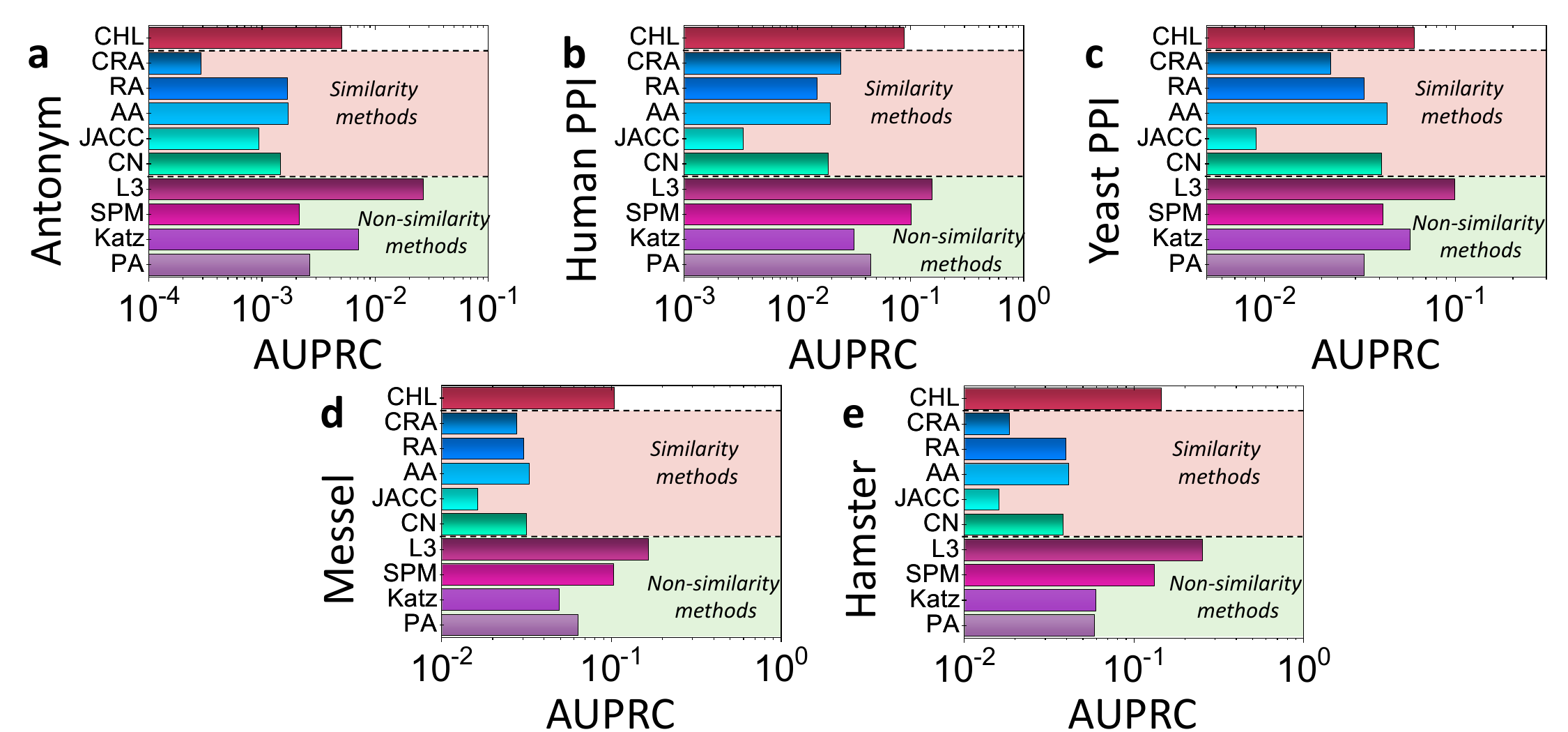}
\caption{\footnotesize (color online) Link prediction results obtained with the Complementarity HyperLink (CHL) are compared to non-embedding link prediction methods. We conduct link prediction experiments in {\bf a} the network of antonyms (Antonyms), {\bf b} human protein-protein interaction network (Human PPI), {\bf c} \emph{S. cerevisiae}  protein-protein interaction network (Yeast PPI), {\bf d} Messel food web (Messel), and ({\bf d}) Hamsterster social network (Hamster) for a fraction $1-q=0.5$ of removed links.  Considered link prediction methods are the (CHL) Complementarity HyperLink (our method), Cannistraci Resource Allocation (CRA)~\cite{Cannistraci2013b}, Resource Allocation (RA)~\cite{Zhou2009predicting}, Adamic Adar (AA)~\cite{Adamic2003friends}, Jaccard Index (JACC)~\cite{Jaccard1901}, the number of Common Neighbors (CN)~\cite{Liben2003link}, the $L3$ method~\cite{kovacs2019network}, the Structural Perturbation Method (SPM)~\cite{lu2015toward}, Katz index~\cite{Katz1953new} with parameter $\beta = 0.1$ (Katz), and Preferential Attachment (PA)~\cite{Barabasi1999}, \label{fig:32}}
\end{figure}

Our link prediction results are summarized in Fig.~\ref{fig:31} and Fig.~\ref{fig:32}. Figure~\ref{fig:31} compares link prediction results of the CHL with other embedding methods. For comparison, we used the ordinary Hyper Link embedder (HL),  Global Vectors for Node Representations (GVNR), Singular Value Decomposition (SVD),  as well as the popular node2vec~\cite{Grover2016node2vec} (N2VEC) and DeepWalk~\cite{perozzi2014deep} (DWALK) network embedding methods.  As seen in Fig.~\ref{fig:31}, link prediction accuracy of the CHL is substantially higher than that of the similarity-based embedding methods HL, N2VEC, and DWALK. The accuracy of the CHL also exceeds that of the GVNR and in three out of five networks that of the SVD.
The high accuracy of the SVD is not surprising as it uses a representation with two vectors per node - a necessary component for the complementarity framework, as we conjectured in Section~\ref{sec:minimal_compl}. Somewhat more surprising is the baseline performance of the GVNR embedding method that is also using a representation with two vectors per node and should have potential at representing complementarity-based systems. A possible reason for the sub-optimal performance of the GVNR is the use of the random walks to enrich the topology of the input network. Random walks may average out the complementarity structure of the input network by enforcing triangle closure.

To get a better picture, we also compared the link prediction accuracy of CHL to representative non-embedding link prediction methods, Fig.~\ref{fig:32}, finding that CHL outperforms all similarity-based methods such as the Resource Allocation (RA)~\cite{Zhou2009predicting} and the Adamic-Adar(AA)~\cite{Adamic2003friends}. This result is expected since similarity-based link prediction methods promote triangle closure, which contradicts the complementarity principle. Also, as expected, we note that all non-similarity based methods, including the baseline preferential attachment (PA) method~\cite{Barabasi1999}, perform better than similarity-based methods on the five selected complementarity-based networks. While CHL is also competitive when compared to non-similarity link prediction methods, we find that its accuracy is at most second best in our experiments, falling behind the $L3$~\cite{kovacs2019network} method and, occasionally, the SPM~\cite{lu2015toward}, or Katz~\cite{Katz1953new} methods.

The competitiveness of complementarity representations in predicting missing links, especially when compared against similarity-based methods, upholds the accuracy of the corresponding representations. While there is room for improvement of the complementarity-based link prediction approach, we leave this problem for future work. Instead, we focus on the representation of the antonym network to gain more insights about its complementarity structure.

\begin{figure}
\includegraphics[width=6.5in]{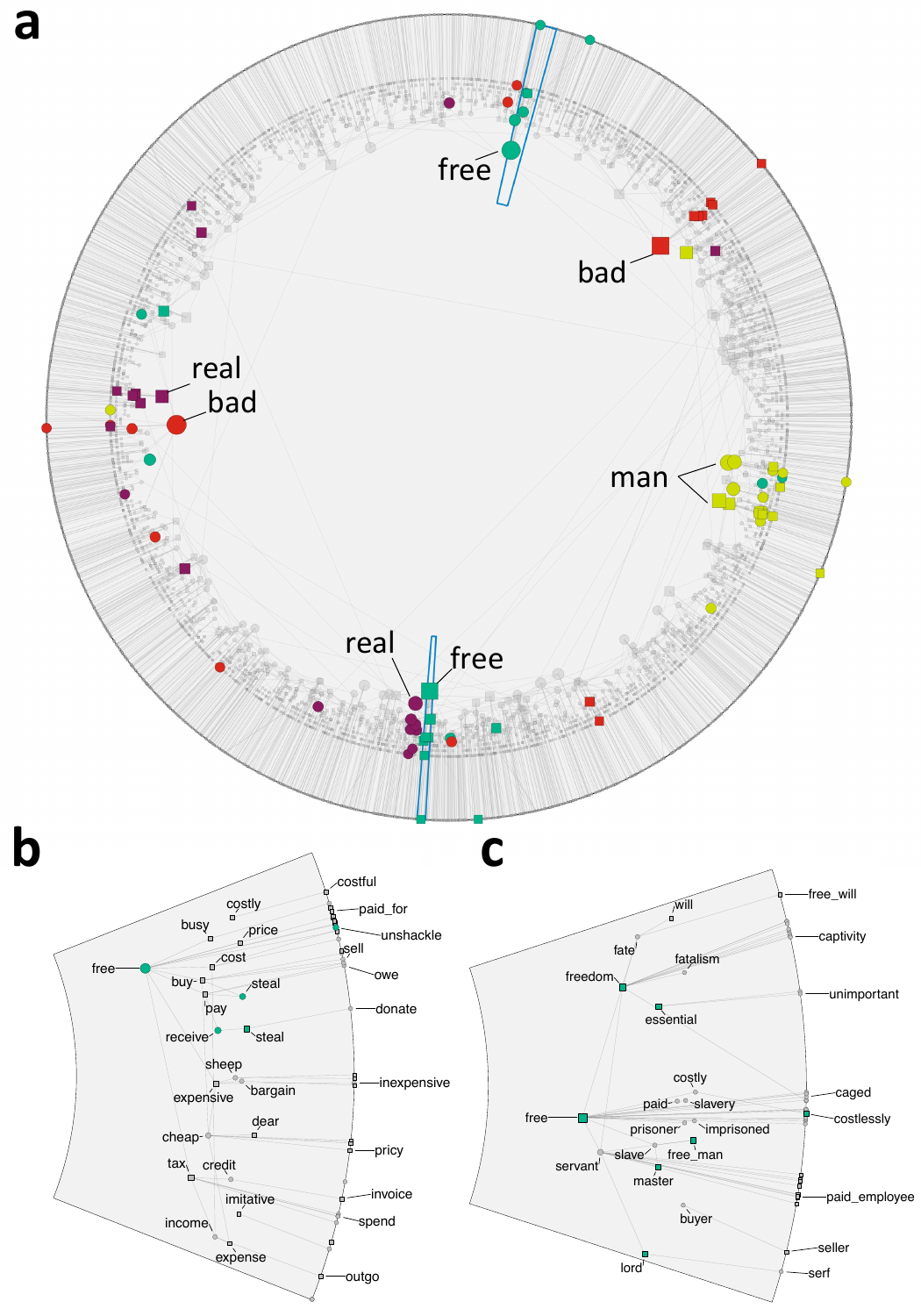}
\caption{\footnotesize (color online) Complementarity representation of the antonym network in a 2-dimensional hyperbolic disk. Each word is represented by two points shown with squares and circles. Larger nodes sizes correspond to more general words, as quantified by the number of their antonyms. Colored are four groups of nodes closest to words Bad, Free, Man, and Real, see Table~\ref{app:synonyms}. To avoid clutter, for every node pair $i$ and $j$ we draw only one link. In doing so, we connect either points $\mathbf{x}_{i}$ and $\mathbf{y}_{j}$ or $\mathbf{x}_{j}$ and $\mathbf{y}_{i}$, selecting the closest pair. {\bf b, c} We zoom into two regions of the space  in the vicinity of the word ``free". We stretch out the angular node coordinates to improve the visibility. }
\label{fig:antonym}
\end{figure}

Figure~\ref{fig:antonym}{\bf a} displays the complementarity representation of the antonym network. While links in the antonym network are established between words with opposite meanings, we can use it to not only find antonyms, as we demonstrated in the link prediction tasks, but also to find synonyms. Indeed, consider two words $i$ and $j$ characterized by points $\left(\mathbf{x}_i, \mathbf{y}_i \right)$ and $\left( \mathbf{x}_j, \mathbf{y}_j \right)$, respectively. While distances between the points of different types -- $d \left[\mathbf{x}_i, \mathbf{y}_j\right]$  and  
$d \left[\mathbf{x}_j, \mathbf{y}_i\right]$ -- quantify complementarity, distances between points of the same type -- $d \left[\mathbf{x}_i, \mathbf{x}_j\right]$ and $d\left[\mathbf{y}_j, \mathbf{y}_i\right]$ -- quantify the similarities, according to the minimal complementarity framework. Using the above intuition, we identified the words closest to the general words ``bad", ``free", ``man", and ``real", finding that most of these words are indeed close in meaning to the seed words, see Table~\ref{app:synonyms}. When marked on the complementarity map of the network, Fig.~\ref{fig:antonym}{\bf a}, each word group is represented by several geometric clusters. In contrast to similarity networks where only one similarity cluster is expected per group of similar nodes, complementarity representations may feature several clusters. This is clearly seen for words in the vicinity of the word ``free", which has two meanings, \emph{no longer confined or imprisoned} and \emph{without cost or payment}. The two distinct meanings correspond to the two points characterizing the word ``free" at 6 and 1 o'clock, respectively, Fig.~\ref{fig:antonym}{\bf b,c}. How many clusters can one expect per group of nodes with similar properties? There can be up to two main clusters in the minimal complementarity framework, each corresponding to a distinct node property. Since each node is represented by two points and only one point is needed to relate the node to other nodes, the remaining point may appear outside the main two clusters.  One example is the word ``escape" that is related to seed word ``free". One of the points characterizing ``escape" is located in the 6 o'clock cluster quantifying the word's similarity with the seed word ``free". The other point corresponding to ``escape" is located around the 9 o'clock on the map, quantifying the word's similarity to words ``away" and ``travel".

\begin{table*}
\begin{tabular}
 {|c|c|c|c|c|}
 \hline \diaghead{zzzzzzzzzzzz}{Rank}{Seed} & Bad & Free & Man & Real \\
\hline \hline 
1&Wrong & Master & Men  & Actual\\
2&Ugly & Freedom & He & Genuine \\
3&Incorrect & Free man & Guy & Truth\\
4&Unpropitious & Essential & Person  & Authentic \\
5&Composite & Escape & Male  & Nothing \\
6&Show & Costlessly & King & Fact\\
7& Uncover & Unshackle & Fellow & Sincere \\
8& Brat & Steal & Wife  & Rigidity \\
9& Mole & Receive & Chick  & Reality \\
10&Beneficial & Lord & Mistress  & Farm \\
 \hline
\end{tabular}
\caption{Seed words (top row) and the words in their hyperbolic vicinity in the CHL representation of the antonym networks. Words are ranked based on similarity distances to the corresponding seed word $d\left[\mathbf{x}_{\rm seed},\mathbf{x}\right]$ and $d\left[ \mathbf{y}_{\rm seed},\mathbf{y}\right]$. }
\label{app:synonyms}
\end{table*}
\begin{figure}
\includegraphics[width=3in]{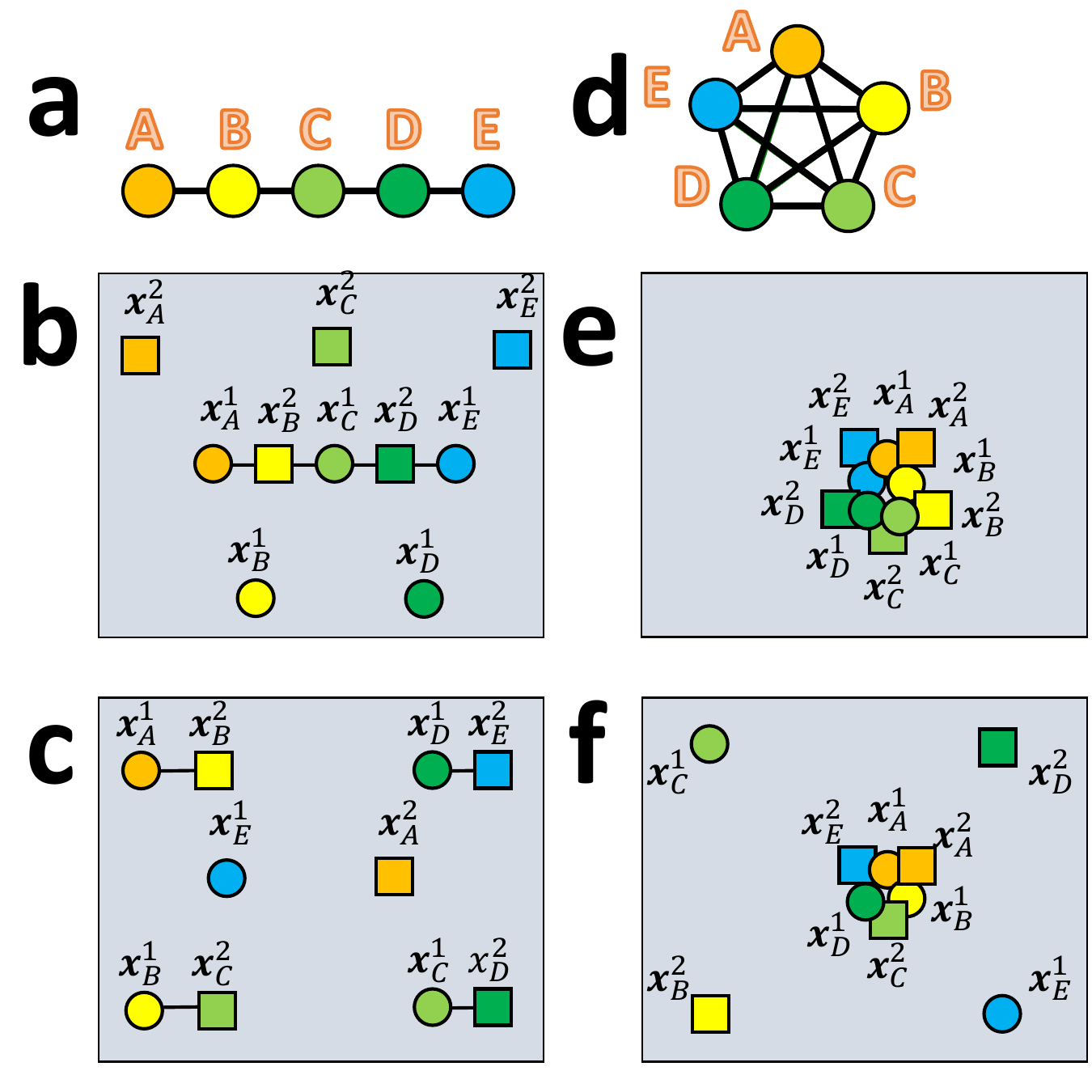}
\caption{\footnotesize (color online) {\bf Paths and communities in complementarity-driven networks.} Shown in ({\bf a}) is the toy network consisting of a chain of $5$ nodes as well as ({\bf b})-({\bf c}) two configurations of points in the space that might lead to it. ({\bf b}) In this configuration the chain of connections observable in the network is achieved by aligning complementary points into a geometric trajectory in the latent space. ({\bf c}) This configuration demonstrates how a chain of connections in ({\bf a}) may arise from a collection of pairwise proximities between points in the latent space. In contrast, trajectory-based alignment is the only possibility for similarity-driven networks. Shown in ({\bf d}) is the toy network consisting of a $5$ node clique as well as ({\bf e})-({\bf f}) two configuration of points in the space that might lead to it. ({\bf e}) In this configuration the clique network  arises due to the complete clustering of all points. ({\bf f}) This configuration demonstrates how the clique network in ({\bf d}) may arise from partially clustered points in the latent space. In contrast, the complete clustering of all points in the latent space is the only possibility for the clique similarity-driven network.}
\label{fig:trajectory}
\end{figure}

\section{Discussion}

We developed two frameworks to quantify complementarity mechanisms in networks. The first principled framework defines complementarity as a phenomenon emerging from the collaboration of agents with related skills. Within the principled framework, collaborating agents bring different but related skills that complement each other in their actions. While this picture is inspired by scientific collaborations and team formation problems, with minimal adjustments it can be translated into domains of systems biology, ecological networks and natural language processing. We anticipate that the principled framework will serve as a foundation for network complementarity models tailored to specific problems and systems of interest. We relied on the principled framework to develop a practical minimal complementarity framework that can be readily used to generate synthetic complementarity-based networks and learn complementarity representations of real networks. We envision that the insights from our work will help us optimize existing and develop novel analysis methods in Science of Science, where interdisciplinary collaborations are of great importance, as well as BioMedicine, which relies on networks of molecular interactions to understand human diseases.

By comparing our complementarity framework to existing learning approaches, we found that Singular Value Decomposition and Global Vectors for Node Representations are potentially applicable to complementarity-driven systems. Our results indicate, however, that additional improvements might required for these methods to optimize their predictive power in complementarity-based systems. 

Since complementarity views each network node as represented by more than one point per node, an important question arises if complementarity can be embedded into a higher-dimensional similarity representation. Indeed, consider a minimal complementarity framework where each node corresponds to two points in a $d$-dimensional latent space. From a formal standpoint, each node is characterized by $2d$ features or vector components and connection probability between the node is a certain function of $4d$ variables. Thus, can we see the complementarity framework as a similarity framework where each node corresponds to a single vector,  not in a $d$-dimensional but $2d$-dimensional latent space? Our current answer is \emph{no}, at least not within the similarity philosophy. According to similarity, the closer two vectors or two nodes are (depending on the representation), the higher the similarity between the two, while maximal similarity is achieved when the two vectors become parallel or two points coincide. According to complementarity, this is not at all necessary: two nodes can be maximally complementary to each other, even if there is a match with respect to one of their vectors or points. 
How do we know if the system at hand is either similarity or complementarity driven? One method to assess the relative presence of the two mechanisms is by measuring the relative densities of triangles and quadrangles in the network of interest~\cite{talaga2022structural}. The authors proposing this method confirm that the majority of studied biological networks are predominantly complementarity driven, while social networks may have a significant presence of both similarity and complementarity, as quantified by densities of triangles and quadrangles, respectively. A more general conclusion can be made, however, that there are no purely similarity-driven or complementarity-driven systems. 
Luckily, our results indicate that similarity may be a special case of complementarity. Our minimal complementarity framework, in particular, contains both complementarity and similarity principles, the latter is captured by the special case where each node is characterized by two identical points in the space.

The complementarity vision not only opens new avenues for the analysis of complementarity-driven systems but also challenges traditional approaches of network science that were initially developed for social networks and are routinely applied to other network classes. Shortest paths and communities have been adopted from similarity-based networks and are routinely used in the analysis of complementarity-driven networks. Network communities are routinely used to quantify disease and functional modules in biological networks~\cite{ghiassian2015disease,ahn2010link,mahmoud2014comm}, and scientific communities in collaboration networks~\cite{Girvan2002a,fortunato2010community}. Shortest paths, on the other hand, are often used to quantify network-based separations between network modules of interest~\cite{menche2015uncovering,Sonawane2019}. 

One example is the notion of the shortest path, which is often envisioned as a certain discrete trajectory in the network space, Fig.~\ref{fig:trajectory}{\bf a}. Such a trajectory is also possible in a complementarity framework: a chain of connections may form due to a spatial alignment of complementary points into a geometric trajectory, Fig.~\ref{fig:trajectory}{\bf b}. While such an alignment is definitely sufficient for the formation of a network chain, it is by no means necessary: another possibility  is a collection of pairwise, yet disjoint, proximities between corresponding points in a latent space, as seen in Fig.~\ref{fig:trajectory}{\bf c}.

Another example is that of the network community. In its classical formulation, a community is a group of nodes densely connected within and sparsely connected outside the group~\cite{Girvan2002a}. Based on this definition, network communities in social sciences are often envisioned as collections of similar node points that are localized in the network, Fig.~\ref{fig:trajectory}{\bf d-f}. In a complementarity driven system, similar nodes and nodes forming a relatively dense subgraph are two distinct concepts since similar nodes are not expected to be connected and, conversely, connected nodes are not expected to be similar. As we demonstrated in the complementarity representation of the antonym network, Fig.~\ref{fig:antonym}{\bf a}, similar nodes may form several localized clusters in the latent space, each cluster corresponding to a different feature. Densely connected subgraphs that are routinely defined as communities in network science, on the other hand, can be either fully, Fig.~\ref{fig:trajectory}{\bf e}, or partially localized in the complementarity representation, Fig.~\ref{fig:trajectory}{\bf f}.

In summary, we hope that the complementarity concepts developed in our work will not only enhance our intuition but also enrich the arsenals of available quantitative methods for the analysis of networks.

{\bf Acknowledgements}

We thank P. Van Mieghem, R. Aldecoa, C. V. Cannistraci, D. Zinoviev, D. Korkin, P. van der Hoorn, D. Krioukov, I. Voitalov, H. Hartle, L. Torres, and B. Klein for useful discussions and suggestions. This work was supported by  ARO Grant No. W911NF-17-1-0491, NSF Grant No. IIS-1741355, the NExTWORKx project, and the Dutch Research Council (NWO) grant OCENW.M20.244.

\newpage
\appendix

\section{Complementarity-based networks}
\label{sec:net_properties}
In our work, we employ five networks where complementarity plays an important role. Here we provide basic description of these networks. Basic topological properties of these networks are described in Table~\ref{app:table}.

The antonym network is the network of words and short phrases, network links are established between words or short phrases if those are antonyms. The network is a subset of semantic relationships of the ConceptNet database~\cite{speer2017conceptnet}.

Human PPI and Yeast PPI are networks of protein-protein interactions. In both networks, nodes are proteins and links represent interactions between them. The Human PPI network is constructed by the Human Reference Protein Interactome Mapping Project (HuRi)~\cite{luck2020reference}. The Yeast PPI network is the network of protein-protein interactions of \emph{S. cerevisiae} commonly known as baker's yeast. We obtain the Yeast PPI network from the BioGrid interaction database~\cite{Stark2006}.

Messel is a foodweb network of the Messel Shale~\cite{dunne2014highly}. The nodes are organisms or taxa and the links are feeding relationships between them. Finally, Hamster or hamsterster is the social network consisting of friendship ties between the users of the website {\it hamsterster.com}. The network is obtained from the KONECT database~\cite{kunegis2013konect}.

\section{Complementarity HyperLink embedder}
\label{app:embedding}

We learn complementarity representations by  repurposing our HyperLink (HL) embedder~\cite{kitsak2020link}. The HL embedder is designed for embedding similarity-driven networks into a 2-dimensional hyperbolic disk $\mathbb{H}^{2}$ using maximum likelihood estimation. In more precise terms, the HL embedder 
aims to find node coordinates in $\mathbb{H}^{2}$ by maximizing the likelihood that the network of interest is generated as a Random Hyperbolic Graph. The HL embedder is freely available at the github~\cite{codeHLembedder}. 

The Complementarity HyperLink (CHL) embedder finds node coordinates in $\mathbb{H}^{2}$ by maximizing the likelihood that the network of interest is generated by the Complementarity Random Hyperbolic Graph (CRHG) model. The CHL embedder is almost identical to the HL embedder save for the function it is maximizing.
To facilitate the embedding of a trimmed network, with links missing at rate  $1-q$ uniformly at random, the CHL embedder is looking for node coordinate assignment $\{\mathbf{x}_{i}\}$ and $\{\mathbf{y}_{i}\}$ for every node $i$ in the network, $i=1,\ldots,N$, maximizing the posterior probability
\begin{equation}
\mathcal{L}\left( \{\mathbf{x}_{i},\mathbf{y}_{i}\} | a_{ij}, \mathcal{P}\right) =
 \frac{\mathcal{L}\left(  a_{ij}| \{\mathbf{x}_{i},\mathbf{y}_{i}\}, \mathcal{P}\right) {\rm P} (\{\mathbf{x}_i\}){\rm P} (\{\mathbf{y}_i\})}{\mathcal{L}\left( a_{ij}| \mathcal{P} \right)},
\label{eq:general_likelihood}
\end{equation}
where $\mathcal{P}$ are the CRHG properties,  $a_{ij}$ are the network adjacency matrix elements, and $\mathcal{L}\left(  a_{ij}| \{\mathbf{x}_{i}, \mathbf{y}_{i}\}\right)$ is given by Eq.~(\ref{eq:likelihood}).  $P(\{\mathbf{x}_i\})$ and $P(\{\mathbf{y}_i \})$ are the prior probabilities of latent coordinates given by the model, which we assume to be independent. We will release the CHL embedder code upon publication of the article.

Other than the posterior probability, the CHL embedder is nearly identical to the HL embedder. As a result, CHL embedder has the same running time complexity cost of $O\left(N^{2}\right)$, where $N$ is the number of nodes.

\begin{figure}
\includegraphics[width=3in]{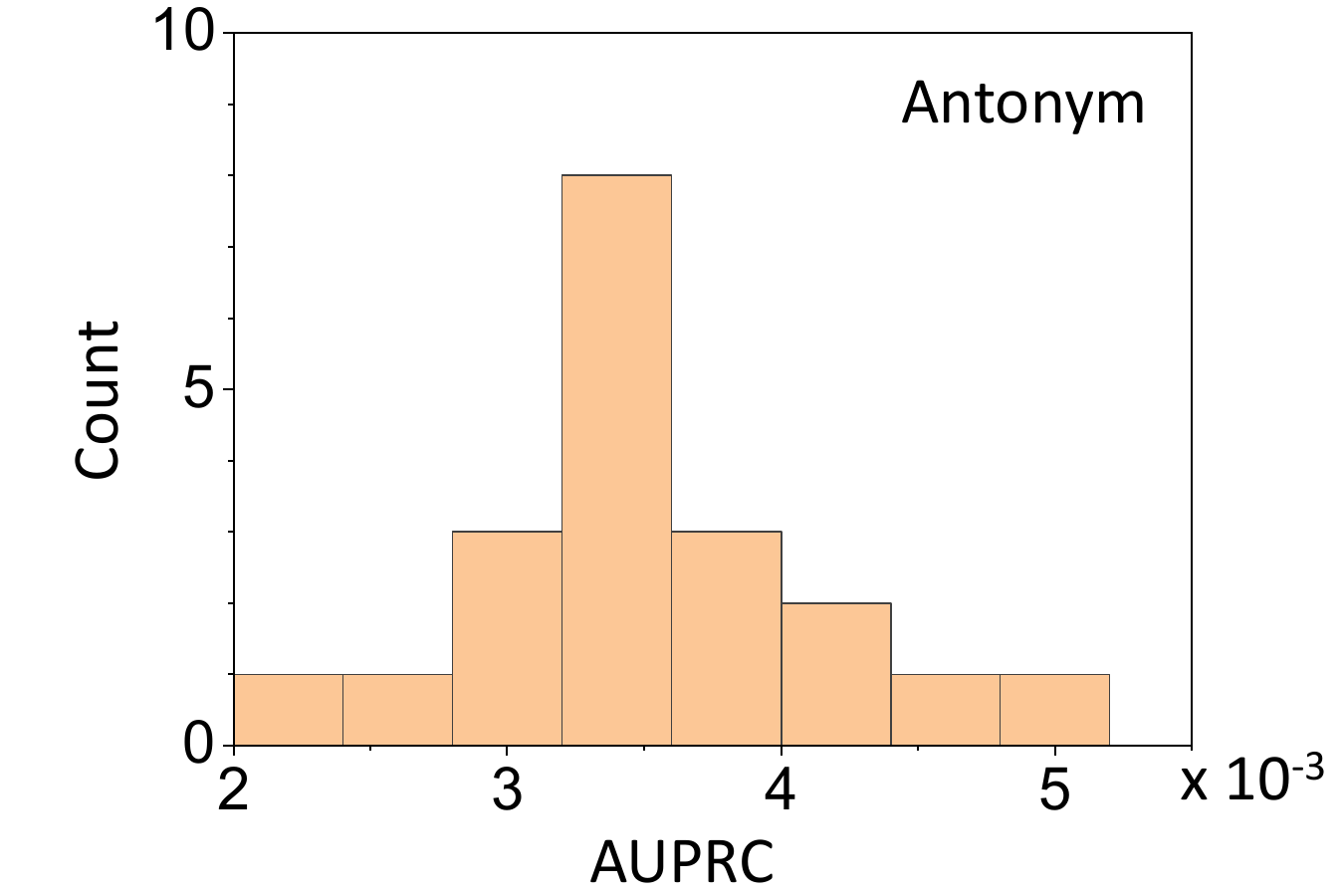}
\caption{\footnotesize (color online) The distribution of link prediction AUPR scores for $20$ independently learned complementarity representations of the Antonym network.}
\label{fig:lp_antonym}
\end{figure}

To test the stability of our complementarity representations, we performed $20$ independent complementarity embeddings of an incomplete complementarity network and evaluated its accuracy in link prediction experiments. As seen from Fig.~\ref{fig:lp_antonym}, all independent complementarity representations results in similar link prediction scores narrowly distributed around the mean of ${\rm AUPR} = 3.5\times 10^{-3}$.

\section{Higher-order generalizations of the complementarity framework}
\label{app:higher_order}

The complementarity framework proposed in the main text assumes that each network node is mapped to two points in the latent space, each point corresponding to a unique feature or characteristic of a node. Here, we sketch a possible generalization of the complementarity framework to the case of an arbitrary number of features per node.

We formulate such a generalization using the terminology of a collaboration network. We assume a collection of nodes, $i = 1,\ldots,N$, each of which is characterized by $M$ different skills, $\{\mathbf{x}^{m}_{i}\}$, $m=1,\ldots,M$.  Any two nodes $i$ and $j$ are connected in the complementarity network if they jointly engage into at least one task $k$, $k = 1,\ldots,K$. Depending on the context, tasks can be either explicit, e.g., scientific publications or problems in a collaboration network, or implicit (unobserved), e.g., biological functions in molecular interaction networks. 

Thus, we introduce a collection of tasks, each of which is characterized by features $\{\mathbf{y}_{k}\}$. The basic building block is the probability $p^{m}\left(\mathbf{x}^{m}_{i}, \mathbf{y}_{k}\right)$ that node $i$ engages in task $k$ using its skill $\mathbf{x}^{m}_{i}$. The probability for node $i$ to engage into task $k$ through any of its tasks independently is
\begin{equation}
q_{ik} = 1 - \prod_{m=1}^{M} \left[1 - p^{m}\left(\mathbf{x}^{m}_{i}, \mathbf{y}_{k}\right)\right],
\end{equation}
and the probability for any two nodes $i$ and $j$ to interact is the probability that any two nodes co-engage in at least one task:
\begin{equation}
p_{ij} = 1 - \prod_{k=1}^{K}(1 - q_{ik}q_{jk}).
\label{eq:general_pij}
\end{equation}
In case $q_{ik} \ll 1$, the connection probability in Eq.~(\ref{eq:general_pij}) can be approximated to the leading order as
\begin{eqnarray}
\label{eq:general_framework1}
p_{ij} &=& \sum_{m=1}^{M}\sum_{n=1}^{M} g^{mn}\left(\mathbf{x}^{m}_{i},\mathbf{x}^{n}_{j}\right),\\
\label{eq:general_framework2}
g^{mn}\left(\mathbf{x}^{m}_{i},\mathbf{x}^{n}_{j}\right) &\equiv& \sum_{k=1}^{K} p^{m}\left(\mathbf{x}^{m}_{i}, \mathbf{y}_{k}\right) p^{n}\left(\mathbf{x}^{n}_{j}, \mathbf{y}_{k}\right)
\end{eqnarray}

As seen from Eq.~(\ref{eq:general_framework2}), the general framework contains both similarity and complementarity components. In case $M=1$, each network node has only one feature and the connection probability $p_{ij}$ depends on the mutual match between features $\mathbf{x}^{1}_i$ and $\mathbf{x}^{1}_j$. The case of $M=2$ contains both similarity, $g^{11}$ and $g^{22}$, and complementarity components, $g^{12}$ and $g^{21}$.

We note that functions $g^{mn}\left(\mathbf{x},\mathbf{y}\right)$ comprising the minimal complementarity  framework in Eqs.~(\ref{eq:general_framework1}) and (\ref{eq:general_framework2}) are not necessarily  geometric, and their functional form should be learned from the network of interest.

It is established that ordinary geometric networks, including random geometric graphs and random hyperbolic graphs, are characterized by a strong clustering coefficient due to the constraints imposed by the triangle inequality. Recent results in Ref.~\cite{krioukov2016clustering} imply that strong clustering should be a sufficient condition for network geometricity. What are the network geometricity conditions for complementarity-driven networks remains an open question.

\begin{table*}
\begin{tabular}
 {|c|c|c|c|}
 \hline Network& Number of nodes, $N$ & Number of links, $E$  & Average degree, $\langle k \rangle$\\
 \hline Antonyms & $5,912$ & $7,986$ & $2.70$\\
 \hline Human PPI & $6,480$ & $31,576$ & $9.75$ \\
 \hline Yeast PPI & $3,488$ & $11,386$  & $6.53$\\
 \hline Messel & $701$ & $6,395$  & $18.25$\\
  \hline Hamster & $1,788$ & $12,476$  & $13,96$ \\
 \hline
\end{tabular}
\caption{Basic properties of largest connected component of the complementarity-based networks studied in this work.}
\label{app:table}
\end{table*}

\section{Complementarity and Bipartite Networks}
\label{app:bipartite}
Several parallels can be drawn between the complementarity-driven networks and bipartite (multipartite) networks.

In a bipartite network, nodes are split into two classes or domains and connections are possible only between the nodes of different classes. The principled complementarity framework operates with agents and tasks: agents that are able to co-solve tasks are complementary to each other. Therefore, synthetic networks generated by the principled complementarity framework are bipartite. 

In a one-mode projection, nodes of one domain of the bipartite network are removed. Nodes of the other domain are connected if they share at least one common neighbor in the bipartite network. Therefore, complementarity links between the agents in the principled complementarity framework can be viewed as one-mode projections of the full network onto the domain of agents. 

Coordinates of agents and tasks in the principled complementarity framework can be viewed as hidden variables, such that connections between the agents and tasks are established independently with probabilities that are functions of the corresponding hidden variables. Therefore, the principled complementarity framework belongs to the class of bipartite networks with hidden variables~\cite{kitsak2011hidden}. 

Furthermore, one can think of a reduced complementarity framework if each agent is represented by one point either in $\mathcal{M}_{1}$ or $\mathcal{M}_{2}$ but not in both. The reduced complementarity framework may be instrumental in modeling, for example, scientific collaborations where each scientist has exactly one domain-specific expertise. If this is the case, the reduced complementarity model also belongs to the class of bipartite networks in latent spaces, that we discussed in Ref.~\cite{kitsak2017latent}.


\begin{thebibliography}{70}%
\makeatletter
\providecommand \@ifxundefined [1]{%
 \@ifx{#1\undefined}
}%
\providecommand \@ifnum [1]{%
 \ifnum #1\expandafter \@firstoftwo
 \else \expandafter \@secondoftwo
 \fi
}%
\providecommand \@ifx [1]{%
 \ifx #1\expandafter \@firstoftwo
 \else \expandafter \@secondoftwo
 \fi
}%
\providecommand \natexlab [1]{#1}%
\providecommand \enquote  [1]{``#1''}%
\providecommand \bibnamefont  [1]{#1}%
\providecommand \bibfnamefont [1]{#1}%
\providecommand \citenamefont [1]{#1}%
\providecommand \href@noop [0]{\@secondoftwo}%
\providecommand \href [0]{\begingroup \@sanitize@url \@href}%
\providecommand \@href[1]{\@@startlink{#1}\@@href}%
\providecommand \@@href[1]{\endgroup#1\@@endlink}%
\providecommand \@sanitize@url [0]{\catcode `\\12\catcode `\$12\catcode
  `\&12\catcode `\#12\catcode `\^12\catcode `\_12\catcode `\%12\relax}%
\providecommand \@@startlink[1]{}%
\providecommand \@@endlink[0]{}%
\providecommand \url  [0]{\begingroup\@sanitize@url \@url }%
\providecommand \@url [1]{\endgroup\@href {#1}{\urlprefix }}%
\providecommand \urlprefix  [0]{URL }%
\providecommand \Eprint [0]{\href }%
\providecommand \doibase [0]{http://dx.doi.org/}%
\providecommand \selectlanguage [0]{\@gobble}%
\providecommand \bibinfo  [0]{\@secondoftwo}%
\providecommand \bibfield  [0]{\@secondoftwo}%
\providecommand \translation [1]{[#1]}%
\providecommand \BibitemOpen [0]{}%
\providecommand \bibitemStop [0]{}%
\providecommand \bibitemNoStop [0]{.\EOS\space}%
\providecommand \EOS [0]{\spacefactor3000\relax}%
\providecommand \BibitemShut  [1]{\csname bibitem#1\endcsname}%
\let\auto@bib@innerbib\@empty
\bibitem [{\citenamefont {McCoy}\ \emph {et~al.}(1997)\citenamefont {McCoy},
  \citenamefont {{Chandana Epa}},\ and\ \citenamefont
  {Colman}}]{mccoy1997electrostatic}%
  \BibitemOpen
  \bibfield  {author} {\bibinfo {author} {\bibfnamefont {A.~J.}\ \bibnamefont
  {McCoy}}, \bibinfo {author} {\bibfnamefont {V.}~\bibnamefont {{Chandana
  Epa}}}, \ and\ \bibinfo {author} {\bibfnamefont {P.~M.}\ \bibnamefont
  {Colman}},\ }\bibfield  {title} {\emph {\bibinfo {title} {{Electrostatic
  complementarity at protein/protein interfaces}},\ }}\href {\doibase
  10.1006/jmbi.1997.0987} {\bibfield  {journal} {\bibinfo  {journal} {J. Mol.
  Biol.}\ }\textbf {\bibinfo {volume} {268}},\ \bibinfo {pages} {570} (\bibinfo
  {year} {1997})}\BibitemShut {NoStop}%
\bibitem [{\citenamefont {Lewis}\ and\ \citenamefont
  {Engelman}(1983)}]{lewis1983lipid}%
  \BibitemOpen
  \bibfield  {author} {\bibinfo {author} {\bibfnamefont {B.~A.}\ \bibnamefont
  {Lewis}}\ and\ \bibinfo {author} {\bibfnamefont {D.~M.}\ \bibnamefont
  {Engelman}},\ }\bibfield  {title} {\emph {\bibinfo {title} {{Lipid bilayer
  thickness varies linearly with acyl chain length in fluid phosphatidylcholine
  vesicles}},\ }}\href {\doibase 10.1016/S0022-2836(83)80007-2} {\bibfield
  {journal} {\bibinfo  {journal} {J. Mol. Biol.}\ }\textbf {\bibinfo {volume}
  {166}},\ \bibinfo {pages} {211} (\bibinfo {year} {1983})}\BibitemShut
  {NoStop}%
\bibitem [{\citenamefont {Yano}\ and\ \citenamefont
  {Matsuzaki}(2006)}]{yano2006measurement}%
  \BibitemOpen
  \bibfield  {author} {\bibinfo {author} {\bibfnamefont {Y.}~\bibnamefont
  {Yano}}\ and\ \bibinfo {author} {\bibfnamefont {K.}~\bibnamefont
  {Matsuzaki}},\ }\bibfield  {title} {\emph {\bibinfo {title} {{Measurement of
  Thermodynamic Parameters for Hydrophobic Mismatch 1: Self-Association of a
  Transmembrane Helix ?}},\ }}\href {\doibase 10.1021/bi0522854} {\bibfield
  {journal} {\bibinfo  {journal} {Biochemistry}\ }\textbf {\bibinfo {volume}
  {45}},\ \bibinfo {pages} {3370} (\bibinfo {year} {2006})}\BibitemShut
  {NoStop}%
\bibitem [{\citenamefont {Botelho}\ \emph {et~al.}(2006)\citenamefont
  {Botelho}, \citenamefont {Huber}, \citenamefont {Sakmar},\ and\ \citenamefont
  {Brown}}]{botelho2006curvature}%
  \BibitemOpen
  \bibfield  {author} {\bibinfo {author} {\bibfnamefont {A.~V.}\ \bibnamefont
  {Botelho}}, \bibinfo {author} {\bibfnamefont {T.}~\bibnamefont {Huber}},
  \bibinfo {author} {\bibfnamefont {T.~P.}\ \bibnamefont {Sakmar}}, \ and\
  \bibinfo {author} {\bibfnamefont {M.~F.}\ \bibnamefont {Brown}},\ }\bibfield
  {title} {\emph {\bibinfo {title} {{Curvature and Hydrophobic Forces Drive
  Oligomerization and Modulate Activity of Rhodopsin in Membranes}},\ }}\href
  {\doibase 10.1529/biophysj.106.082776} {\bibfield  {journal} {\bibinfo
  {journal} {Biophys. J.}\ }\textbf {\bibinfo {volume} {91}},\ \bibinfo {pages}
  {4464} (\bibinfo {year} {2006})}\BibitemShut {NoStop}%
\bibitem [{\citenamefont {Zhang}\ \emph {et~al.}(2009)\citenamefont {Zhang},
  \citenamefont {Sanner},\ and\ \citenamefont {Olson}}]{zhang2009shape}%
  \BibitemOpen
  \bibfield  {author} {\bibinfo {author} {\bibfnamefont {Q.}~\bibnamefont
  {Zhang}}, \bibinfo {author} {\bibfnamefont {M.}~\bibnamefont {Sanner}}, \
  and\ \bibinfo {author} {\bibfnamefont {A.~J.}\ \bibnamefont {Olson}},\
  }\bibfield  {title} {\emph {\bibinfo {title} {{Shape complementarity of
  protein-protein complexes at multiple resolutions}},\ }}\href {\doibase
  10.1002/prot.22256} {\bibfield  {journal} {\bibinfo  {journal} {Proteins
  Struct. Funct. Bioinforma.}\ }\textbf {\bibinfo {volume} {75}},\ \bibinfo
  {pages} {453} (\bibinfo {year} {2009})}\BibitemShut {NoStop}%
\bibitem [{\citenamefont {Li}\ \emph {et~al.}(2013)\citenamefont {Li},
  \citenamefont {Zhang},\ and\ \citenamefont {Cao}}]{li2013role}%
  \BibitemOpen
  \bibfield  {author} {\bibinfo {author} {\bibfnamefont {Y.}~\bibnamefont
  {Li}}, \bibinfo {author} {\bibfnamefont {X.}~\bibnamefont {Zhang}}, \ and\
  \bibinfo {author} {\bibfnamefont {D.}~\bibnamefont {Cao}},\ }\bibfield
  {title} {\emph {\bibinfo {title} {{The Role of Shape Complementarity in the
  Protein-Protein Interactions}},\ }}\href {\doibase 10.1038/srep03271}
  {\bibfield  {journal} {\bibinfo  {journal} {Sci. Rep.}\ }\textbf {\bibinfo
  {volume} {3}},\ \bibinfo {pages} {3271} (\bibinfo {year} {2013})}\BibitemShut
  {NoStop}%
\bibitem [{\citenamefont {Mattsson}\ \emph {et~al.}(2021)\citenamefont
  {Mattsson}, \citenamefont {Takes}, \citenamefont {Heemskerk}, \citenamefont
  {Diks}, \citenamefont {Buiten}, \citenamefont {Faber},\ and\ \citenamefont
  {Sloot}}]{mattsson2021functional}%
  \BibitemOpen
  \bibfield  {author} {\bibinfo {author} {\bibfnamefont {C.~E.~S.}\
  \bibnamefont {Mattsson}}, \bibinfo {author} {\bibfnamefont {F.~W.}\
  \bibnamefont {Takes}}, \bibinfo {author} {\bibfnamefont {E.~M.}\ \bibnamefont
  {Heemskerk}}, \bibinfo {author} {\bibfnamefont {C.}~\bibnamefont {Diks}},
  \bibinfo {author} {\bibfnamefont {G.}~\bibnamefont {Buiten}}, \bibinfo
  {author} {\bibfnamefont {A.}~\bibnamefont {Faber}}, \ and\ \bibinfo {author}
  {\bibfnamefont {P.~M.~A.}\ \bibnamefont {Sloot}},\ }\bibfield  {title} {\emph
  {\bibinfo {title} {{Functional Structure in Production Networks}},\ }}\href
  {\doibase 10.3389/fdata.2021.666712} {\bibfield  {journal} {\bibinfo
  {journal} {Front. Big Data}\ }\textbf {\bibinfo {volume} {4}} (\bibinfo
  {year} {2021}),\ 10.3389/fdata.2021.666712},\ \Eprint
  {http://arxiv.org/abs/2103.15777} {arXiv:2103.15777} \BibitemShut {NoStop}%
\bibitem [{\citenamefont {McFarland}\ and\ \citenamefont
  {Brown}(1973)}]{mcfarland1973social}%
  \BibitemOpen
  \bibfield  {author} {\bibinfo {author} {\bibfnamefont {D.~D.}\ \bibnamefont
  {McFarland}}\ and\ \bibinfo {author} {\bibfnamefont {D.~J.}\ \bibnamefont
  {Brown}},\ }\bibfield  {title} {\emph {\bibinfo {title} {{Social distance as
  a metric: a systematic introduction to smallest space analysis}},\
  }}\href@noop {} {\bibfield  {journal} {\bibinfo  {journal} {Bond. Plur. Form
  Subst. Urban Soc. Networks}\ }\textbf {\bibinfo {volume} {6}},\ \bibinfo
  {pages} {213} (\bibinfo {year} {1973})}\BibitemShut {NoStop}%
\bibitem [{\citenamefont {Cannistraci}\ \emph
  {et~al.}(2013{\natexlab{a}})\citenamefont {Cannistraci}, \citenamefont
  {Alanis-Lobato},\ and\ \citenamefont {Ravasi}}]{cannistraci2013minimum}%
  \BibitemOpen
  \bibfield  {author} {\bibinfo {author} {\bibfnamefont {C.~V.}\ \bibnamefont
  {Cannistraci}}, \bibinfo {author} {\bibfnamefont {G.}~\bibnamefont
  {Alanis-Lobato}}, \ and\ \bibinfo {author} {\bibfnamefont {T.}~\bibnamefont
  {Ravasi}},\ }\bibfield  {title} {\emph {\bibinfo {title} {{Minimum
  curvilinearity to enhance topological prediction of protein interactions by
  network embedding}},\ }}\href {\doibase 10.1093/bioinformatics/btt208}
  {\bibfield  {journal} {\bibinfo  {journal} {Bioinformatics}\ }\textbf
  {\bibinfo {volume} {29}},\ \bibinfo {pages} {i199} (\bibinfo {year}
  {2013}{\natexlab{a}})}\BibitemShut {NoStop}%
\bibitem [{\citenamefont {Yang}\ \emph {et~al.}(2014)\citenamefont {Yang},
  \citenamefont {Yih}, \citenamefont {He}, \citenamefont {Gao},\ and\
  \citenamefont {Deng}}]{yang2015embedding}%
  \BibitemOpen
  \bibfield  {author} {\bibinfo {author} {\bibfnamefont {B.}~\bibnamefont
  {Yang}}, \bibinfo {author} {\bibfnamefont {W.-t.}\ \bibnamefont {Yih}},
  \bibinfo {author} {\bibfnamefont {X.}~\bibnamefont {He}}, \bibinfo {author}
  {\bibfnamefont {J.}~\bibnamefont {Gao}}, \ and\ \bibinfo {author}
  {\bibfnamefont {L.}~\bibnamefont {Deng}},\ }\bibfield  {title} {\emph
  {\bibinfo {title} {{Embedding Entities and Relations for Learning and
  Inference in Knowledge Bases}},\ }}\href {http://arxiv.org/abs/1412.6575}
  {\bibfield  {journal} {\bibinfo  {journal} {3rd Int. Conf. Learn. Represent.
  ICLR 2015 - Conf. Track Proc.}\ } (\bibinfo {year} {2014})},\ \Eprint
  {http://arxiv.org/abs/1412.6575} {arXiv:1412.6575} \BibitemShut {NoStop}%
\bibitem [{\citenamefont {Xiao}\ \emph {et~al.}(2015)\citenamefont {Xiao},
  \citenamefont {Huang},\ and\ \citenamefont {Zhu}}]{xiao2015from}%
  \BibitemOpen
  \bibfield  {author} {\bibinfo {author} {\bibfnamefont {H.}~\bibnamefont
  {Xiao}}, \bibinfo {author} {\bibfnamefont {M.}~\bibnamefont {Huang}}, \ and\
  \bibinfo {author} {\bibfnamefont {X.}~\bibnamefont {Zhu}},\ }\bibfield
  {title} {\emph {\bibinfo {title} {{From One Point to A Manifold: Knowledge
  Graph Embedding For Precise Link Prediction}},\ }}\href
  {http://arxiv.org/abs/1512.04792} {\bibfield  {journal} {\bibinfo  {journal}
  {IJCAI Int. Jt. Conf. Artif. Intell.}\ } (\bibinfo {year} {2015})},\ \Eprint
  {http://arxiv.org/abs/1512.04792} {arXiv:1512.04792} \BibitemShut {NoStop}%
\bibitem [{\citenamefont {Tang}\ \emph {et~al.}(2015)\citenamefont {Tang},
  \citenamefont {Qu}, \citenamefont {Wang}, \citenamefont {Zhang},
  \citenamefont {Yan},\ and\ \citenamefont {Mei}}]{tang2015line}%
  \BibitemOpen
  \bibfield  {author} {\bibinfo {author} {\bibfnamefont {J.}~\bibnamefont
  {Tang}}, \bibinfo {author} {\bibfnamefont {M.}~\bibnamefont {Qu}}, \bibinfo
  {author} {\bibfnamefont {M.}~\bibnamefont {Wang}}, \bibinfo {author}
  {\bibfnamefont {M.}~\bibnamefont {Zhang}}, \bibinfo {author} {\bibfnamefont
  {J.}~\bibnamefont {Yan}}, \ and\ \bibinfo {author} {\bibfnamefont
  {Q.}~\bibnamefont {Mei}},\ }in\ \href {\doibase 10.1145/2736277.2741093}
  {\emph {\bibinfo {booktitle} {Proc. 24th Int. Conf. World Wide Web}}}\
  (\bibinfo  {publisher} {International World Wide Web Conferences Steering
  Committee},\ \bibinfo {address} {Republic and Canton of Geneva,
  Switzerland},\ \bibinfo {year} {2015})\ pp.\ \bibinfo {pages}
  {1067--1077}\BibitemShut {NoStop}%
\bibitem [{\citenamefont {Grover}\ and\ \citenamefont
  {Leskovec}(2016)}]{Grover2016node2vec}%
  \BibitemOpen
  \bibfield  {author} {\bibinfo {author} {\bibfnamefont {A.}~\bibnamefont
  {Grover}}\ and\ \bibinfo {author} {\bibfnamefont {J.}~\bibnamefont
  {Leskovec}},\ }in\ \href {\doibase 10.1145/2939672.2939754} {\emph {\bibinfo
  {booktitle} {Proc. 22nd ACM SIGKDD Int. Conf. Knowl. Discov. Data Min.}}}\
  (\bibinfo  {publisher} {ACM},\ \bibinfo {address} {New York, NY, USA},\
  \bibinfo {year} {2016})\ pp.\ \bibinfo {pages} {855--864},\ \Eprint
  {http://arxiv.org/abs/1607.00653} {arXiv:1607.00653} \BibitemShut {NoStop}%
\bibitem [{\citenamefont {Zhu}\ \emph {et~al.}(2016)\citenamefont {Zhu},
  \citenamefont {Guo}, \citenamefont {Yin}, \citenamefont {Steeg},\ and\
  \citenamefont {Galstyan}}]{zhu2016scalable}%
  \BibitemOpen
  \bibfield  {author} {\bibinfo {author} {\bibfnamefont {L.}~\bibnamefont
  {Zhu}}, \bibinfo {author} {\bibfnamefont {D.}~\bibnamefont {Guo}}, \bibinfo
  {author} {\bibfnamefont {J.}~\bibnamefont {Yin}}, \bibinfo {author}
  {\bibfnamefont {G.~V.}\ \bibnamefont {Steeg}}, \ and\ \bibinfo {author}
  {\bibfnamefont {A.}~\bibnamefont {Galstyan}},\ }\bibfield  {title} {\emph
  {\bibinfo {title} {{Scalable Temporal Latent Space Inference for Link
  Prediction in Dynamic Social Networks}},\ }}\href {\doibase
  10.1109/TKDE.2016.2591009} {\bibfield  {journal} {\bibinfo  {journal} {IEEE
  Trans. Knowl. Data Eng.}\ }\textbf {\bibinfo {volume} {28}},\ \bibinfo
  {pages} {2765} (\bibinfo {year} {2016})}\BibitemShut {NoStop}%
\bibitem [{\citenamefont {Nickel}\ and\ \citenamefont
  {Kiela}(2018)}]{nickel2018learning}%
  \BibitemOpen
  \bibfield  {author} {\bibinfo {author} {\bibfnamefont {M.}~\bibnamefont
  {Nickel}}\ and\ \bibinfo {author} {\bibfnamefont {D.}~\bibnamefont {Kiela}},\
  }in\ \href@noop {} {\emph {\bibinfo {booktitle} {35th Int. Conf. Mach. Learn.
  ICML 2018}}}\ (\bibinfo {year} {2018})\ \Eprint
  {http://arxiv.org/abs/1806.03417} {arXiv:1806.03417} \BibitemShut {NoStop}%
\bibitem [{\citenamefont {Kazemi}\ and\ \citenamefont
  {Poole}(2018)}]{Kazemi2018SimplE}%
  \BibitemOpen
  \bibfield  {author} {\bibinfo {author} {\bibfnamefont {S.~M.}\ \bibnamefont
  {Kazemi}}\ and\ \bibinfo {author} {\bibfnamefont {D.}~\bibnamefont {Poole}},\
  }\bibfield  {title} {\emph {\bibinfo {title} {{SimplE Embedding for Link
  Prediction in Knowledge Graphs}},\ }}\href {http://arxiv.org/abs/1802.04868}
  {\bibfield  {journal} {\bibinfo  {journal} {Adv. Neural Inf. Process. Syst.}\
  } (\bibinfo {year} {2018})},\ \Eprint {http://arxiv.org/abs/1802.04868}
  {arXiv:1802.04868} \BibitemShut {NoStop}%
\bibitem [{\citenamefont {{Brew, Anthony;
  Salter-Townshend}}(2010)}]{brew2010latent}%
  \BibitemOpen
  \bibfield  {author} {\bibinfo {author} {\bibfnamefont {M.}~\bibnamefont
  {{Brew, Anthony; Salter-Townshend}}},\ }\bibfield  {title} {\emph {\bibinfo
  {title} {{A latent space mapping for link prediction}},\ }}\href@noop {}
  {\bibfield  {journal} {\bibinfo  {journal} {Neural Inf. Process.}\ ,\
  \bibinfo {pages} {1}} (\bibinfo {year} {2010})}\BibitemShut {NoStop}%
\bibitem [{\citenamefont {Kitsak}\ \emph {et~al.}(2020)\citenamefont {Kitsak},
  \citenamefont {Voitalov},\ and\ \citenamefont {Krioukov}}]{kitsak2020link}%
  \BibitemOpen
  \bibfield  {author} {\bibinfo {author} {\bibfnamefont {M.}~\bibnamefont
  {Kitsak}}, \bibinfo {author} {\bibfnamefont {I.}~\bibnamefont {Voitalov}}, \
  and\ \bibinfo {author} {\bibfnamefont {D.}~\bibnamefont {Krioukov}},\
  }\bibfield  {title} {\emph {\bibinfo {title} {{Link prediction with
  hyperbolic geometry}},\ }}\href {\doibase 10.1103/PhysRevResearch.2.043113}
  {\bibfield  {journal} {\bibinfo  {journal} {Phys. Rev. Res.}\ }\textbf
  {\bibinfo {volume} {2}},\ \bibinfo {pages} {043113} (\bibinfo {year}
  {2020})}\BibitemShut {NoStop}%
\bibitem [{\citenamefont {Garc{\'{i}}a-P{\'{e}}rez}\ \emph
  {et~al.}(2020)\citenamefont {Garc{\'{i}}a-P{\'{e}}rez}, \citenamefont
  {Aliakbarisani}, \citenamefont {Ghasemi},\ and\ \citenamefont
  {Serrano}}]{perez2020precision}%
  \BibitemOpen
  \bibfield  {author} {\bibinfo {author} {\bibfnamefont {G.}~\bibnamefont
  {Garc{\'{i}}a-P{\'{e}}rez}}, \bibinfo {author} {\bibfnamefont
  {R.}~\bibnamefont {Aliakbarisani}}, \bibinfo {author} {\bibfnamefont
  {A.}~\bibnamefont {Ghasemi}}, \ and\ \bibinfo {author} {\bibfnamefont
  {M.~{\'{A}}.}\ \bibnamefont {Serrano}},\ }\bibfield  {title} {\emph {\bibinfo
  {title} {{Precision as a measure of predictability of missing links in real
  networks}},\ }}\href {\doibase 10.1103/PhysRevE.101.052318} {\bibfield
  {journal} {\bibinfo  {journal} {Phys. Rev. E}\ }\textbf {\bibinfo {volume}
  {101}},\ \bibinfo {pages} {052318} (\bibinfo {year} {2020})}\BibitemShut
  {NoStop}%
\bibitem [{\citenamefont {Newman}\ and\ \citenamefont
  {Peixoto}(1984)}]{newman2015generalized}%
  \BibitemOpen
  \bibfield  {author} {\bibinfo {author} {\bibfnamefont {M.~E.~J.}\
  \bibnamefont {Newman}}\ and\ \bibinfo {author} {\bibfnamefont {T.~P.}\
  \bibnamefont {Peixoto}},\ }\bibfield  {title} {\emph {\bibinfo {title}
  {{Generalized communities in networks}},\ }}\href {\doibase
  10.1103/PhysRevLett.115.088701} {\bibfield  {journal} {\bibinfo  {journal}
  {Phys. Rev. Lett.}\ }\textbf {\bibinfo {volume} {115}},\ \bibinfo {pages}
  {88701} (\bibinfo {year} {1984})}\BibitemShut {NoStop}%
\bibitem [{\citenamefont {Zuev}\ \emph {et~al.}(2015)\citenamefont {Zuev},
  \citenamefont {Bogu{\~{n}}{\'{a}}}, \citenamefont {Bianconi},\ and\
  \citenamefont {Krioukov}}]{zuev2015emergence}%
  \BibitemOpen
  \bibfield  {author} {\bibinfo {author} {\bibfnamefont {K.}~\bibnamefont
  {Zuev}}, \bibinfo {author} {\bibfnamefont {M.}~\bibnamefont
  {Bogu{\~{n}}{\'{a}}}}, \bibinfo {author} {\bibfnamefont {G.}~\bibnamefont
  {Bianconi}}, \ and\ \bibinfo {author} {\bibfnamefont {D.}~\bibnamefont
  {Krioukov}},\ }\bibfield  {title} {\emph {\bibinfo {title} {{Emergence of
  Soft Communities from Geometric Preferential Attachment}},\ }}\href {\doibase
  10.1038/srep09421} {\bibfield  {journal} {\bibinfo  {journal} {Sci. Rep.}\
  }\textbf {\bibinfo {volume} {5}},\ \bibinfo {pages} {9421} (\bibinfo {year}
  {2015})}\BibitemShut {NoStop}%
\bibitem [{\citenamefont {Yang}\ \emph {et~al.}(2015)\citenamefont {Yang},
  \citenamefont {Cao}, \citenamefont {Jin}, \citenamefont {Wang},\ and\
  \citenamefont {Meng}}]{Yang2015unified}%
  \BibitemOpen
  \bibfield  {author} {\bibinfo {author} {\bibfnamefont {L.}~\bibnamefont
  {Yang}}, \bibinfo {author} {\bibfnamefont {X.}~\bibnamefont {Cao}}, \bibinfo
  {author} {\bibfnamefont {D.}~\bibnamefont {Jin}}, \bibinfo {author}
  {\bibfnamefont {X.}~\bibnamefont {Wang}}, \ and\ \bibinfo {author}
  {\bibfnamefont {D.}~\bibnamefont {Meng}},\ }\bibfield  {title} {\emph
  {\bibinfo {title} {{A Unified Semi-Supervised Community Detection Framework
  Using Latent Space Graph Regularization}},\ }}\href {\doibase
  10.1109/TCYB.2014.2377154} {\bibfield  {journal} {\bibinfo  {journal} {IEEE
  Trans. Cybern.}\ }\textbf {\bibinfo {volume} {45}},\ \bibinfo {pages} {2585}
  (\bibinfo {year} {2015})}\BibitemShut {NoStop}%
\bibitem [{\citenamefont {Sewell}\ and\ \citenamefont
  {Chen}(2017)}]{Sewell2017latent}%
  \BibitemOpen
  \bibfield  {author} {\bibinfo {author} {\bibfnamefont {D.~K.}\ \bibnamefont
  {Sewell}}\ and\ \bibinfo {author} {\bibfnamefont {Y.}~\bibnamefont {Chen}},\
  }\bibfield  {title} {\emph {\bibinfo {title} {{Latent Space Approaches to
  Community Detection in Dynamic Networks}},\ }}\href {\doibase
  10.1214/16-BA1000} {\bibfield  {journal} {\bibinfo  {journal} {Bayesian
  Anal.}\ }\textbf {\bibinfo {volume} {12}} (\bibinfo {year} {2017}),\
  10.1214/16-BA1000}\BibitemShut {NoStop}%
\bibitem [{\citenamefont {Bogu{\~{n}}{\'{a}}}\ \emph
  {et~al.}(2010)\citenamefont {Bogu{\~{n}}{\'{a}}}, \citenamefont
  {Papadopoulos},\ and\ \citenamefont {Krioukov}}]{Boguna2010sustaining}%
  \BibitemOpen
  \bibfield  {author} {\bibinfo {author} {\bibfnamefont {M.}~\bibnamefont
  {Bogu{\~{n}}{\'{a}}}}, \bibinfo {author} {\bibfnamefont {F.}~\bibnamefont
  {Papadopoulos}}, \ and\ \bibinfo {author} {\bibfnamefont {D.}~\bibnamefont
  {Krioukov}},\ }\bibfield  {title} {\emph {\bibinfo {title} {{Sustaining the
  Internet with hyperbolic mapping}},\ }}\href {\doibase 10.1038/ncomms1063}
  {\bibfield  {journal} {\bibinfo  {journal} {Nat. Commun.}\ }\textbf {\bibinfo
  {volume} {1}},\ \bibinfo {pages} {62} (\bibinfo {year} {2010})}\BibitemShut
  {NoStop}%
\bibitem [{\citenamefont {Guly{\'{a}}s}\ \emph {et~al.}(2015)\citenamefont
  {Guly{\'{a}}s}, \citenamefont {B{\'{i}}r{\'{o}}}, \citenamefont
  {K?r{\"{o}}si}, \citenamefont {R{\'{e}}tv{\'{a}}ri},\ and\ \citenamefont
  {Krioukov}}]{Gulyas2015}%
  \BibitemOpen
  \bibfield  {author} {\bibinfo {author} {\bibfnamefont {A.}~\bibnamefont
  {Guly{\'{a}}s}}, \bibinfo {author} {\bibfnamefont {J.~J.}\ \bibnamefont
  {B{\'{i}}r{\'{o}}}}, \bibinfo {author} {\bibfnamefont {A.}~\bibnamefont
  {K\H{o}r{\"{o}}si}}, \bibinfo {author} {\bibfnamefont {G.}~\bibnamefont
  {R{\'{e}}tv{\'{a}}ri}}, \ and\ \bibinfo {author} {\bibfnamefont
  {D.}~\bibnamefont {Krioukov}},\ }\bibfield  {title} {\emph {\bibinfo {title}
  {{Navigable networks as Nash equilibria of navigation games}},\ }}\href
  {\doibase 10.1038/ncomms8651} {\bibfield  {journal} {\bibinfo  {journal}
  {Nat. Commun.}\ }\textbf {\bibinfo {volume} {6}},\ \bibinfo {pages} {7651}
  (\bibinfo {year} {2015})}\BibitemShut {NoStop}%
\bibitem [{\citenamefont {Voitalov}\ \emph {et~al.}(2017)\citenamefont
  {Voitalov}, \citenamefont {Aldecoa}, \citenamefont {Wang},\ and\
  \citenamefont {Krioukov}}]{voitalov2017geohyperbolic}%
  \BibitemOpen
  \bibfield  {author} {\bibinfo {author} {\bibfnamefont {I.}~\bibnamefont
  {Voitalov}}, \bibinfo {author} {\bibfnamefont {R.}~\bibnamefont {Aldecoa}},
  \bibinfo {author} {\bibfnamefont {L.}~\bibnamefont {Wang}}, \ and\ \bibinfo
  {author} {\bibfnamefont {D.}~\bibnamefont {Krioukov}},\ }\bibfield  {title}
  {\emph {\bibinfo {title} {{Geohyperbolic Routing and Addressing Schemes}},\
  }}\href {\doibase 10.1145/3138808.3138811} {\bibfield  {journal} {\bibinfo
  {journal} {ACM SIGCOMM Comput. Commun. Rev.}\ }\textbf {\bibinfo {volume}
  {47}},\ \bibinfo {pages} {11} (\bibinfo {year} {2017})}\BibitemShut {NoStop}%
\bibitem [{\citenamefont {Garc{\'{i}}a-P{\'{e}}rez}\ \emph
  {et~al.}(2018)\citenamefont {Garc{\'{i}}a-P{\'{e}}rez}, \citenamefont
  {Bogu{\~{n}}{\'{a}}},\ and\ \citenamefont {Serrano}}]{Garcia2018multiscale}%
  \BibitemOpen
  \bibfield  {author} {\bibinfo {author} {\bibfnamefont {G.}~\bibnamefont
  {Garc{\'{i}}a-P{\'{e}}rez}}, \bibinfo {author} {\bibfnamefont
  {M.}~\bibnamefont {Bogu{\~{n}}{\'{a}}}}, \ and\ \bibinfo {author}
  {\bibfnamefont {M.~{\'{A}}.}\ \bibnamefont {Serrano}},\ }\bibfield  {title}
  {\emph {\bibinfo {title} {{Multiscale unfolding of real networks by geometric
  renormalization}},\ }}\href {\doibase 10.1038/s41567-018-0072-5} {\bibfield
  {journal} {\bibinfo  {journal} {Nat. Phys.}\ }\textbf {\bibinfo {volume}
  {14}},\ \bibinfo {pages} {583} (\bibinfo {year} {2018})}\BibitemShut
  {NoStop}%
\bibitem [{\citenamefont {Kleinberg}(2006)}]{kleinberg2006complex}%
  \BibitemOpen
  \bibfield  {author} {\bibinfo {author} {\bibfnamefont {J.}~\bibnamefont
  {Kleinberg}},\ }in\ \href {\doibase 10.4171/022-3/50} {\emph {\bibinfo
  {booktitle} {Proc. Int. Congr. Math. Madrid, August 22?30, 2006}}}\
  (\bibinfo  {publisher} {European Mathematical Society Publishing House},\
  \bibinfo {address} {Zuerich, Switzerland},\ \bibinfo {year} {2006})\ pp.\
  \bibinfo {pages} {1019--1044}\BibitemShut {NoStop}%
\bibitem [{\citenamefont {Ratnasamy}\ \emph {et~al.}(2001)\citenamefont
  {Ratnasamy}, \citenamefont {Francis}, \citenamefont {Handley}, \citenamefont
  {Karp},\ and\ \citenamefont {Shenker}}]{ratnasamy2001scalable}%
  \BibitemOpen
  \bibfield  {author} {\bibinfo {author} {\bibfnamefont {S.}~\bibnamefont
  {Ratnasamy}}, \bibinfo {author} {\bibfnamefont {P.}~\bibnamefont {Francis}},
  \bibinfo {author} {\bibfnamefont {M.}~\bibnamefont {Handley}}, \bibinfo
  {author} {\bibfnamefont {R.}~\bibnamefont {Karp}}, \ and\ \bibinfo {author}
  {\bibfnamefont {S.}~\bibnamefont {Shenker}},\ }\bibfield  {title} {\emph
  {\bibinfo {title} {{A scalable content-addressable network}},\ }}\href
  {\doibase 10.1145/964723.383072} {\bibfield  {journal} {\bibinfo  {journal}
  {ACM SIGCOMM Comput. Commun. Rev.}\ }\textbf {\bibinfo {volume} {31}},\
  \bibinfo {pages} {161} (\bibinfo {year} {2001})}\BibitemShut {NoStop}%
\bibitem [{\citenamefont {Risson}\ and\ \citenamefont
  {Moors}(2006)}]{risson2006survey}%
  \BibitemOpen
  \bibfield  {author} {\bibinfo {author} {\bibfnamefont {J.}~\bibnamefont
  {Risson}}\ and\ \bibinfo {author} {\bibfnamefont {T.}~\bibnamefont {Moors}},\
  }\bibfield  {title} {\emph {\bibinfo {title} {{Survey of research towards
  robust peer-to-peer networks: Search methods}},\ }}\href {\doibase
  10.1016/j.comnet.2006.02.001} {\bibfield  {journal} {\bibinfo  {journal}
  {Comput. Networks}\ }\textbf {\bibinfo {volume} {50}},\ \bibinfo {pages}
  {3485} (\bibinfo {year} {2006})}\BibitemShut {NoStop}%
\bibitem [{\citenamefont {Kov{\'{a}}cs}\ \emph {et~al.}(2019)\citenamefont
  {Kov{\'{a}}cs}, \citenamefont {Luck}, \citenamefont {Spirohn}, \citenamefont
  {Wang}, \citenamefont {Pollis}, \citenamefont {Schlabach}, \citenamefont
  {Bian}, \citenamefont {Kim}, \citenamefont {Kishore}, \citenamefont {Hao},
  \citenamefont {Calderwood}, \citenamefont {Vidal},\ and\ \citenamefont
  {Barab{\'{a}}si}}]{kovacs2019network}%
  \BibitemOpen
  \bibfield  {author} {\bibinfo {author} {\bibfnamefont {I.~A.}\ \bibnamefont
  {Kov{\'{a}}cs}}, \bibinfo {author} {\bibfnamefont {K.}~\bibnamefont {Luck}},
  \bibinfo {author} {\bibfnamefont {K.}~\bibnamefont {Spirohn}}, \bibinfo
  {author} {\bibfnamefont {Y.}~\bibnamefont {Wang}}, \bibinfo {author}
  {\bibfnamefont {C.}~\bibnamefont {Pollis}}, \bibinfo {author} {\bibfnamefont
  {S.}~\bibnamefont {Schlabach}}, \bibinfo {author} {\bibfnamefont
  {W.}~\bibnamefont {Bian}}, \bibinfo {author} {\bibfnamefont {D.-K.}\
  \bibnamefont {Kim}}, \bibinfo {author} {\bibfnamefont {N.}~\bibnamefont
  {Kishore}}, \bibinfo {author} {\bibfnamefont {T.}~\bibnamefont {Hao}},
  \bibinfo {author} {\bibfnamefont {M.~A.}\ \bibnamefont {Calderwood}},
  \bibinfo {author} {\bibfnamefont {M.}~\bibnamefont {Vidal}}, \ and\ \bibinfo
  {author} {\bibfnamefont {A.-L.}\ \bibnamefont {Barab{\'{a}}si}},\ }\bibfield
  {title} {\emph {\bibinfo {title} {{Network-based prediction of protein
  interactions}},\ }}\href {\doibase 10.1038/s41467-019-09177-y} {\bibfield
  {journal} {\bibinfo  {journal} {Nat. Commun.}\ }\textbf {\bibinfo {volume}
  {10}},\ \bibinfo {pages} {1240} (\bibinfo {year} {2019})}\BibitemShut
  {NoStop}%
\bibitem [{\citenamefont {Talaga}\ and\ \citenamefont
  {Nowak}(2022)}]{talaga2022structural}%
  \BibitemOpen
  \bibfield  {author} {\bibinfo {author} {\bibfnamefont {S.}~\bibnamefont
  {Talaga}}\ and\ \bibinfo {author} {\bibfnamefont {A.}~\bibnamefont {Nowak}},\
  }\bibfield  {title} {\emph {\bibinfo {title} {{Structural measures of
  similarity and complementarity in complex networks}},\ }}\href {\doibase
  10.1038/s41598-022-20710-w} {\bibfield  {journal} {\bibinfo  {journal} {Sci.
  Rep.}\ }\textbf {\bibinfo {volume} {12}},\ \bibinfo {pages} {16580} (\bibinfo
  {year} {2022})}\BibitemShut {NoStop}%
\bibitem [{\citenamefont {Bogu{\~{n}}{\'{a}}}\ and\ \citenamefont
  {Pastor-Satorras}(2003)}]{boguna2003class}%
  \BibitemOpen
  \bibfield  {author} {\bibinfo {author} {\bibfnamefont {M.}~\bibnamefont
  {Bogu{\~{n}}{\'{a}}}}\ and\ \bibinfo {author} {\bibfnamefont
  {R.}~\bibnamefont {Pastor-Satorras}},\ }\bibfield  {title} {\emph {\bibinfo
  {title} {{Class of correlated random networks with hidden variables}},\
  }}\href {\doibase 10.1103/PhysRevE.68.036112} {\bibfield  {journal} {\bibinfo
   {journal} {Phys. Rev. E}\ }\textbf {\bibinfo {volume} {68}},\ \bibinfo
  {pages} {036112} (\bibinfo {year} {2003})}\BibitemShut {NoStop}%
\bibitem [{\citenamefont {Kitsak}\ and\ \citenamefont
  {Krioukov}(2011)}]{kitsak2011hidden}%
  \BibitemOpen
  \bibfield  {author} {\bibinfo {author} {\bibfnamefont {M.}~\bibnamefont
  {Kitsak}}\ and\ \bibinfo {author} {\bibfnamefont {D.}~\bibnamefont
  {Krioukov}},\ }\bibfield  {title} {\emph {\bibinfo {title} {{Hidden variables
  in bipartite networks}},\ }}\href {\doibase 10.1103/PhysRevE.84.026114}
  {\bibfield  {journal} {\bibinfo  {journal} {Phys. Rev. E}\ }\textbf {\bibinfo
  {volume} {84}},\ \bibinfo {pages} {026114} (\bibinfo {year}
  {2011})}\BibitemShut {NoStop}%
\bibitem [{\citenamefont {Sinatra}\ \emph {et~al.}(2015)\citenamefont
  {Sinatra}, \citenamefont {Deville}, \citenamefont {Szell}, \citenamefont
  {Wang},\ and\ \citenamefont {Barab{\'{a}}si}}]{Sinatra2015century}%
  \BibitemOpen
  \bibfield  {author} {\bibinfo {author} {\bibfnamefont {R.}~\bibnamefont
  {Sinatra}}, \bibinfo {author} {\bibfnamefont {P.}~\bibnamefont {Deville}},
  \bibinfo {author} {\bibfnamefont {M.}~\bibnamefont {Szell}}, \bibinfo
  {author} {\bibfnamefont {D.}~\bibnamefont {Wang}}, \ and\ \bibinfo {author}
  {\bibfnamefont {A.-L.}\ \bibnamefont {Barab{\'{a}}si}},\ }\bibfield  {title}
  {\emph {\bibinfo {title} {{A century of physics}},\ }}\href {\doibase
  10.1038/nphys3494} {\bibfield  {journal} {\bibinfo  {journal} {Nat. Phys.}\
  }\textbf {\bibinfo {volume} {11}},\ \bibinfo {pages} {791} (\bibinfo {year}
  {2015})}\BibitemShut {NoStop}%
\bibitem [{\citenamefont {Pan}\ \emph {et~al.}(2012)\citenamefont {Pan},
  \citenamefont {Sinha}, \citenamefont {Kaski},\ and\ \citenamefont
  {Saram{\"{a}}ki}}]{Pan2012evolution}%
  \BibitemOpen
  \bibfield  {author} {\bibinfo {author} {\bibfnamefont {R.~K.}\ \bibnamefont
  {Pan}}, \bibinfo {author} {\bibfnamefont {S.}~\bibnamefont {Sinha}}, \bibinfo
  {author} {\bibfnamefont {K.}~\bibnamefont {Kaski}}, \ and\ \bibinfo {author}
  {\bibfnamefont {J.}~\bibnamefont {Saram{\"{a}}ki}},\ }\bibfield  {title}
  {\emph {\bibinfo {title} {{The evolution of interdisciplinarity in physics
  research}},\ }}\href {\doibase 10.1038/srep00551} {\bibfield  {journal}
  {\bibinfo  {journal} {Sci. Rep.}\ }\textbf {\bibinfo {volume} {2}},\ \bibinfo
  {pages} {551} (\bibinfo {year} {2012})}\BibitemShut {NoStop}%
\bibitem [{\citenamefont {Gates}\ \emph {et~al.}(2019)\citenamefont {Gates},
  \citenamefont {Ke}, \citenamefont {Varol},\ and\ \citenamefont
  {Barab{\'{a}}si}}]{Gates2019nature}%
  \BibitemOpen
  \bibfield  {author} {\bibinfo {author} {\bibfnamefont {A.~J.}\ \bibnamefont
  {Gates}}, \bibinfo {author} {\bibfnamefont {Q.}~\bibnamefont {Ke}}, \bibinfo
  {author} {\bibfnamefont {O.}~\bibnamefont {Varol}}, \ and\ \bibinfo {author}
  {\bibfnamefont {A.-L.}\ \bibnamefont {Barab{\'{a}}si}},\ }\bibfield  {title}
  {\emph {\bibinfo {title} {{Nature's reach: narrow work has broad impact}},\
  }}\href {\doibase 10.1038/d41586-019-03308-7} {\bibfield  {journal} {\bibinfo
   {journal} {Nature}\ }\textbf {\bibinfo {volume} {575}},\ \bibinfo {pages}
  {32} (\bibinfo {year} {2019})}\BibitemShut {NoStop}%
\bibitem [{\citenamefont {Pennington}\ \emph {et~al.}(2014)\citenamefont
  {Pennington}, \citenamefont {Socher},\ and\ \citenamefont
  {Manning}}]{pennington2014glove}%
  \BibitemOpen
  \bibfield  {author} {\bibinfo {author} {\bibfnamefont {J.}~\bibnamefont
  {Pennington}}, \bibinfo {author} {\bibfnamefont {R.}~\bibnamefont {Socher}},
  \ and\ \bibinfo {author} {\bibfnamefont {C.}~\bibnamefont {Manning}},\ }in\
  \href {\doibase 10.3115/v1/D14-1162} {\emph {\bibinfo {booktitle} {Proc. 2014
  Conf. Empir. Methods Nat. Lang. Process.}}}\ (\bibinfo  {publisher}
  {Association for Computational Linguistics},\ \bibinfo {address}
  {Stroudsburg, PA, USA},\ \bibinfo {year} {2014})\ pp.\ \bibinfo {pages}
  {1532--1543}\BibitemShut {NoStop}%
\bibitem [{\citenamefont {Brochier}\ \emph {et~al.}(2019)\citenamefont
  {Brochier}, \citenamefont {Guille},\ and\ \citenamefont
  {Velcin}}]{brochier2019global}%
  \BibitemOpen
  \bibfield  {author} {\bibinfo {author} {\bibfnamefont {R.}~\bibnamefont
  {Brochier}}, \bibinfo {author} {\bibfnamefont {A.}~\bibnamefont {Guille}}, \
  and\ \bibinfo {author} {\bibfnamefont {J.}~\bibnamefont {Velcin}},\ }in\
  \href {\doibase 10.1145/3308558.3313595} {\emph {\bibinfo {booktitle} {World
  Wide Web Conf.}}}\ (\bibinfo  {publisher} {ACM},\ \bibinfo {address} {New
  York, NY, USA},\ \bibinfo {year} {2019})\ pp.\ \bibinfo {pages}
  {2587--2593}\BibitemShut {NoStop}%
\bibitem [{\citenamefont {Krioukov}\ \emph {et~al.}(2010)\citenamefont
  {Krioukov}, \citenamefont {Papadopoulos}, \citenamefont {Kitsak},
  \citenamefont {Vahdat},\ and\ \citenamefont
  {Bogu{\~{n}}{\'{a}}}}]{Krioukov2010hyperbolic}%
  \BibitemOpen
  \bibfield  {author} {\bibinfo {author} {\bibfnamefont {D.}~\bibnamefont
  {Krioukov}}, \bibinfo {author} {\bibfnamefont {F.}~\bibnamefont
  {Papadopoulos}}, \bibinfo {author} {\bibfnamefont {M.}~\bibnamefont
  {Kitsak}}, \bibinfo {author} {\bibfnamefont {A.}~\bibnamefont {Vahdat}}, \
  and\ \bibinfo {author} {\bibfnamefont {M.}~\bibnamefont
  {Bogu{\~{n}}{\'{a}}}},\ }\bibfield  {title} {\emph {\bibinfo {title}
  {{Hyperbolic geometry of complex networks}},\ }}\href {\doibase
  10.1103/PhysRevE.82.036106} {\bibfield  {journal} {\bibinfo  {journal} {Phys.
  Rev. E}\ }\textbf {\bibinfo {volume} {82}},\ \bibinfo {pages} {036106}
  (\bibinfo {year} {2010})}\BibitemShut {NoStop}%
\bibitem [{\citenamefont {Serrano}\ \emph {et~al.}(2008)\citenamefont
  {Serrano}, \citenamefont {Krioukov},\ and\ \citenamefont
  {Bogu{\~{n}}{\'{a}}}}]{Serrano2008}%
  \BibitemOpen
  \bibfield  {author} {\bibinfo {author} {\bibfnamefont {M.~{\'{A}}.}\
  \bibnamefont {Serrano}}, \bibinfo {author} {\bibfnamefont {D.}~\bibnamefont
  {Krioukov}}, \ and\ \bibinfo {author} {\bibfnamefont {M.}~\bibnamefont
  {Bogu{\~{n}}{\'{a}}}},\ }\bibfield  {title} {\emph {\bibinfo {title}
  {{Self-Similarity of Complex Networks and Hidden Metric Spaces}},\ }}\href
  {\doibase 10.1103/PhysRevLett.100.078701} {\bibfield  {journal} {\bibinfo
  {journal} {Phys. Rev. Lett.}\ }\textbf {\bibinfo {volume} {100}},\ \bibinfo
  {pages} {078701} (\bibinfo {year} {2008})}\BibitemShut {NoStop}%
\bibitem [{\citenamefont {Papadopoulos}\ \emph {et~al.}(2012)\citenamefont
  {Papadopoulos}, \citenamefont {Kitsak}, \citenamefont {Serrano},
  \citenamefont {Bogu{\~{n}}{\'{a}}},\ and\ \citenamefont
  {Krioukov}}]{Papadopoulos2012popularity}%
  \BibitemOpen
  \bibfield  {author} {\bibinfo {author} {\bibfnamefont {F.}~\bibnamefont
  {Papadopoulos}}, \bibinfo {author} {\bibfnamefont {M.}~\bibnamefont
  {Kitsak}}, \bibinfo {author} {\bibfnamefont {M.~{\'{A}}.}\ \bibnamefont
  {Serrano}}, \bibinfo {author} {\bibfnamefont {M.}~\bibnamefont
  {Bogu{\~{n}}{\'{a}}}}, \ and\ \bibinfo {author} {\bibfnamefont
  {D.}~\bibnamefont {Krioukov}},\ }\bibfield  {title} {\emph {\bibinfo {title}
  {{Popularity versus similarity in growing networks}},\ }}\href {\doibase
  10.1038/nature11459} {\bibfield  {journal} {\bibinfo  {journal} {Nature}\
  }\textbf {\bibinfo {volume} {489}},\ \bibinfo {pages} {537} (\bibinfo {year}
  {2012})}\BibitemShut {NoStop}%
\bibitem [{\citenamefont {Iamnitchi}\ \emph {et~al.}(2004)\citenamefont
  {Iamnitchi}, \citenamefont {Ripeanu},\ and\ \citenamefont
  {Foster}}]{iamnitchi2004small}%
  \BibitemOpen
  \bibfield  {author} {\bibinfo {author} {\bibfnamefont {A.}~\bibnamefont
  {Iamnitchi}}, \bibinfo {author} {\bibfnamefont {M.}~\bibnamefont {Ripeanu}},
  \ and\ \bibinfo {author} {\bibfnamefont {I.}~\bibnamefont {Foster}},\ }in\
  \href {\doibase 10.1109/INFCOM.2004.1356982} {\emph {\bibinfo {booktitle}
  {IEEE INFOCOM 2004}}},\ Vol.~\bibinfo {volume} {2},\ \bibinfo {organization}
  {IEEE}\ (\bibinfo  {publisher} {IEEE},\ \bibinfo {year} {2004})\ pp.\
  \bibinfo {pages} {952--963}\BibitemShut {NoStop}%
\bibitem [{\citenamefont {Burgos}\ \emph {et~al.}(2008)\citenamefont {Burgos},
  \citenamefont {Ceva}, \citenamefont {Hern{\'{a}}ndez}, \citenamefont
  {Perazzo}, \citenamefont {Devoto},\ and\ \citenamefont
  {Medan}}]{burgos2008two}%
  \BibitemOpen
  \bibfield  {author} {\bibinfo {author} {\bibfnamefont {E.}~\bibnamefont
  {Burgos}}, \bibinfo {author} {\bibfnamefont {H.}~\bibnamefont {Ceva}},
  \bibinfo {author} {\bibfnamefont {L.}~\bibnamefont {Hern{\'{a}}ndez}},
  \bibinfo {author} {\bibfnamefont {R.~P.~J.}\ \bibnamefont {Perazzo}},
  \bibinfo {author} {\bibfnamefont {M.}~\bibnamefont {Devoto}}, \ and\ \bibinfo
  {author} {\bibfnamefont {D.}~\bibnamefont {Medan}},\ }\bibfield  {title}
  {\emph {\bibinfo {title} {{Two classes of bipartite networks: Nested
  biological and social systems}},\ }}\href {\doibase
  10.1103/PhysRevE.78.046113} {\bibfield  {journal} {\bibinfo  {journal} {Phys.
  Rev. E}\ }\textbf {\bibinfo {volume} {78}},\ \bibinfo {pages} {046113}
  (\bibinfo {year} {2008})}\BibitemShut {NoStop}%
\bibitem [{\citenamefont {Kitsak}\ \emph {et~al.}(2017)\citenamefont {Kitsak},
  \citenamefont {Papadopoulos},\ and\ \citenamefont
  {Krioukov}}]{kitsak2017latent}%
  \BibitemOpen
  \bibfield  {author} {\bibinfo {author} {\bibfnamefont {M.}~\bibnamefont
  {Kitsak}}, \bibinfo {author} {\bibfnamefont {F.}~\bibnamefont
  {Papadopoulos}}, \ and\ \bibinfo {author} {\bibfnamefont {D.}~\bibnamefont
  {Krioukov}},\ }\bibfield  {title} {\emph {\bibinfo {title} {{Latent geometry
  of bipartite networks}},\ }}\href {\doibase 10.1103/PhysRevE.95.032309}
  {\bibfield  {journal} {\bibinfo  {journal} {Phys. Rev. E}\ }\textbf {\bibinfo
  {volume} {95}},\ \bibinfo {pages} {032309} (\bibinfo {year}
  {2017})}\BibitemShut {NoStop}%
\bibitem [{\citenamefont {Das}\ and\ \citenamefont {Yu}(2012)}]{Das2012}%
  \BibitemOpen
  \bibfield  {author} {\bibinfo {author} {\bibfnamefont {J.}~\bibnamefont
  {Das}}\ and\ \bibinfo {author} {\bibfnamefont {H.}~\bibnamefont {Yu}},\
  }\bibfield  {title} {\emph {\bibinfo {title} {{HINT: High-quality protein
  interactomes and their applications in understanding human disease}},\
  }}\href {\doibase 10.1186/1752-0509-6-92} {\bibfield  {journal} {\bibinfo
  {journal} {BMC Syst. Biol.}\ }\textbf {\bibinfo {volume} {6}},\ \bibinfo
  {pages} {92} (\bibinfo {year} {2012})}\BibitemShut {NoStop}%
\bibitem [{\citenamefont {Luck}\ \emph {et~al.}(2020)\citenamefont {Luck},
  \citenamefont {Kim}, \citenamefont {Lambourne}, \citenamefont {Spirohn},
  \citenamefont {Begg}, \citenamefont {Bian}, \citenamefont {Brignall},
  \citenamefont {Cafarelli}, \citenamefont {Campos-Laborie}, \citenamefont
  {Charloteaux}, \citenamefont {Choi}, \citenamefont {Cot{\'{e}}},
  \citenamefont {Daley}, \citenamefont {Deimling}, \citenamefont {Desbuleux},
  \citenamefont {Dricot}, \citenamefont {Gebbia}, \citenamefont {Hardy},
  \citenamefont {Kishore}, \citenamefont {Knapp}, \citenamefont {Kov{\'{a}}cs},
  \citenamefont {Lemmens}, \citenamefont {Mee}, \citenamefont {Mellor},
  \citenamefont {Pollis}, \citenamefont {Pons}, \citenamefont {Richardson},
  \citenamefont {Schlabach}, \citenamefont {Teeking}, \citenamefont {Yadav},
  \citenamefont {Babor}, \citenamefont {Balcha}, \citenamefont {Basha},
  \citenamefont {Bowman-Colin}, \citenamefont {Chin}, \citenamefont {Choi},
  \citenamefont {Colabella}, \citenamefont {Coppin}, \citenamefont {D'Amata},
  \citenamefont {{De Ridder}}, \citenamefont {{De Rouck}}, \citenamefont
  {Duran-Frigola}, \citenamefont {Ennajdaoui}, \citenamefont {Goebels},
  \citenamefont {Goehring}, \citenamefont {Gopal}, \citenamefont {Haddad},
  \citenamefont {Hatchi}, \citenamefont {Helmy}, \citenamefont {Jacob},
  \citenamefont {Kassa}, \citenamefont {Landini}, \citenamefont {Li},
  \citenamefont {van Lieshout}, \citenamefont {MacWilliams}, \citenamefont
  {Markey}, \citenamefont {Paulson}, \citenamefont {Rangarajan}, \citenamefont
  {Rasla}, \citenamefont {Rayhan}, \citenamefont {Rolland}, \citenamefont
  {San-Miguel}, \citenamefont {Shen}, \citenamefont {Sheykhkarimli},
  \citenamefont {Sheynkman}, \citenamefont {Simonovsky}, \citenamefont
  {Ta?an}, \citenamefont {Tejeda}, \citenamefont {Tropepe}, \citenamefont
  {Twizere}, \citenamefont {Wang}, \citenamefont {Weatheritt}, \citenamefont
  {Weile}, \citenamefont {Xia}, \citenamefont {Yang}, \citenamefont
  {Yeger-Lotem}, \citenamefont {Zhong}, \citenamefont {Aloy}, \citenamefont
  {Bader}, \citenamefont {{De Las Rivas}}, \citenamefont {Gaudet},
  \citenamefont {Hao}, \citenamefont {Rak}, \citenamefont {Tavernier},
  \citenamefont {Hill}, \citenamefont {Vidal}, \citenamefont {Roth},\ and\
  \citenamefont {Calderwood}}]{luck2020reference}%
  \BibitemOpen
  \bibfield  {author} {\bibinfo {author} {\bibfnamefont {K.}~\bibnamefont
  {Luck}}, \bibinfo {author} {\bibfnamefont {D.-K.}\ \bibnamefont {Kim}},
  \bibinfo {author} {\bibfnamefont {L.}~\bibnamefont {Lambourne}}, \bibinfo
  {author} {\bibfnamefont {K.}~\bibnamefont {Spirohn}}, \bibinfo {author}
  {\bibfnamefont {B.~E.}\ \bibnamefont {Begg}}, \bibinfo {author}
  {\bibfnamefont {W.}~\bibnamefont {Bian}}, \bibinfo {author} {\bibfnamefont
  {R.}~\bibnamefont {Brignall}}, \bibinfo {author} {\bibfnamefont
  {T.}~\bibnamefont {Cafarelli}}, \bibinfo {author} {\bibfnamefont {F.~J.}\
  \bibnamefont {Campos-Laborie}}, \bibinfo {author} {\bibfnamefont
  {B.}~\bibnamefont {Charloteaux}}, \bibinfo {author} {\bibfnamefont
  {D.}~\bibnamefont {Choi}}, \bibinfo {author} {\bibfnamefont {A.~G.}\
  \bibnamefont {Cot{\'{e}}}}, \bibinfo {author} {\bibfnamefont
  {M.}~\bibnamefont {Daley}}, \bibinfo {author} {\bibfnamefont
  {S.}~\bibnamefont {Deimling}}, \bibinfo {author} {\bibfnamefont
  {A.}~\bibnamefont {Desbuleux}}, \bibinfo {author} {\bibfnamefont
  {A.}~\bibnamefont {Dricot}}, \bibinfo {author} {\bibfnamefont
  {M.}~\bibnamefont {Gebbia}}, \bibinfo {author} {\bibfnamefont {M.~F.}\
  \bibnamefont {Hardy}}, \bibinfo {author} {\bibfnamefont {N.}~\bibnamefont
  {Kishore}}, \bibinfo {author} {\bibfnamefont {J.~J.}\ \bibnamefont {Knapp}},
  \bibinfo {author} {\bibfnamefont {I.~A.}\ \bibnamefont {Kov{\'{a}}cs}},
  \bibinfo {author} {\bibfnamefont {I.}~\bibnamefont {Lemmens}}, \bibinfo
  {author} {\bibfnamefont {M.~W.}\ \bibnamefont {Mee}}, \bibinfo {author}
  {\bibfnamefont {J.~C.}\ \bibnamefont {Mellor}}, \bibinfo {author}
  {\bibfnamefont {C.}~\bibnamefont {Pollis}}, \bibinfo {author} {\bibfnamefont
  {C.}~\bibnamefont {Pons}}, \bibinfo {author} {\bibfnamefont {A.~D.}\
  \bibnamefont {Richardson}}, \bibinfo {author} {\bibfnamefont
  {S.}~\bibnamefont {Schlabach}}, \bibinfo {author} {\bibfnamefont
  {B.}~\bibnamefont {Teeking}}, \bibinfo {author} {\bibfnamefont
  {A.}~\bibnamefont {Yadav}}, \bibinfo {author} {\bibfnamefont
  {M.}~\bibnamefont {Babor}}, \bibinfo {author} {\bibfnamefont
  {D.}~\bibnamefont {Balcha}}, \bibinfo {author} {\bibfnamefont
  {O.}~\bibnamefont {Basha}}, \bibinfo {author} {\bibfnamefont
  {C.}~\bibnamefont {Bowman-Colin}}, \bibinfo {author} {\bibfnamefont {S.-F.}\
  \bibnamefont {Chin}}, \bibinfo {author} {\bibfnamefont {S.~G.}\ \bibnamefont
  {Choi}}, \bibinfo {author} {\bibfnamefont {C.}~\bibnamefont {Colabella}},
  \bibinfo {author} {\bibfnamefont {G.}~\bibnamefont {Coppin}}, \bibinfo
  {author} {\bibfnamefont {C.}~\bibnamefont {D'Amata}}, \bibinfo {author}
  {\bibfnamefont {D.}~\bibnamefont {{De Ridder}}}, \bibinfo {author}
  {\bibfnamefont {S.}~\bibnamefont {{De Rouck}}}, \bibinfo {author}
  {\bibfnamefont {M.}~\bibnamefont {Duran-Frigola}}, \bibinfo {author}
  {\bibfnamefont {H.}~\bibnamefont {Ennajdaoui}}, \bibinfo {author}
  {\bibfnamefont {F.}~\bibnamefont {Goebels}}, \bibinfo {author} {\bibfnamefont
  {L.}~\bibnamefont {Goehring}}, \bibinfo {author} {\bibfnamefont
  {A.}~\bibnamefont {Gopal}}, \bibinfo {author} {\bibfnamefont
  {G.}~\bibnamefont {Haddad}}, \bibinfo {author} {\bibfnamefont
  {E.}~\bibnamefont {Hatchi}}, \bibinfo {author} {\bibfnamefont
  {M.}~\bibnamefont {Helmy}}, \bibinfo {author} {\bibfnamefont
  {Y.}~\bibnamefont {Jacob}}, \bibinfo {author} {\bibfnamefont
  {Y.}~\bibnamefont {Kassa}}, \bibinfo {author} {\bibfnamefont
  {S.}~\bibnamefont {Landini}}, \bibinfo {author} {\bibfnamefont
  {R.}~\bibnamefont {Li}}, \bibinfo {author} {\bibfnamefont {N.}~\bibnamefont
  {van Lieshout}}, \bibinfo {author} {\bibfnamefont {A.}~\bibnamefont
  {MacWilliams}}, \bibinfo {author} {\bibfnamefont {D.}~\bibnamefont {Markey}},
  \bibinfo {author} {\bibfnamefont {J.~N.}\ \bibnamefont {Paulson}}, \bibinfo
  {author} {\bibfnamefont {S.}~\bibnamefont {Rangarajan}}, \bibinfo {author}
  {\bibfnamefont {J.}~\bibnamefont {Rasla}}, \bibinfo {author} {\bibfnamefont
  {A.}~\bibnamefont {Rayhan}}, \bibinfo {author} {\bibfnamefont
  {T.}~\bibnamefont {Rolland}}, \bibinfo {author} {\bibfnamefont
  {A.}~\bibnamefont {San-Miguel}}, \bibinfo {author} {\bibfnamefont
  {Y.}~\bibnamefont {Shen}}, \bibinfo {author} {\bibfnamefont {D.}~\bibnamefont
  {Sheykhkarimli}}, \bibinfo {author} {\bibfnamefont {G.~M.}\ \bibnamefont
  {Sheynkman}}, \bibinfo {author} {\bibfnamefont {E.}~\bibnamefont
  {Simonovsky}}, \bibinfo {author} {\bibfnamefont {M.}~\bibnamefont {Ta\c{s}an}},
  \bibinfo {author} {\bibfnamefont {A.}~\bibnamefont {Tejeda}}, \bibinfo
  {author} {\bibfnamefont {V.}~\bibnamefont {Tropepe}}, \bibinfo {author}
  {\bibfnamefont {J.-C.}\ \bibnamefont {Twizere}}, \bibinfo {author}
  {\bibfnamefont {Y.}~\bibnamefont {Wang}}, \bibinfo {author} {\bibfnamefont
  {R.~J.}\ \bibnamefont {Weatheritt}}, \bibinfo {author} {\bibfnamefont
  {J.}~\bibnamefont {Weile}}, \bibinfo {author} {\bibfnamefont
  {Y.}~\bibnamefont {Xia}}, \bibinfo {author} {\bibfnamefont {X.}~\bibnamefont
  {Yang}}, \bibinfo {author} {\bibfnamefont {E.}~\bibnamefont {Yeger-Lotem}},
  \bibinfo {author} {\bibfnamefont {Q.}~\bibnamefont {Zhong}}, \bibinfo
  {author} {\bibfnamefont {P.}~\bibnamefont {Aloy}}, \bibinfo {author}
  {\bibfnamefont {G.~D.}\ \bibnamefont {Bader}}, \bibinfo {author}
  {\bibfnamefont {J.}~\bibnamefont {{De Las Rivas}}}, \bibinfo {author}
  {\bibfnamefont {S.}~\bibnamefont {Gaudet}}, \bibinfo {author} {\bibfnamefont
  {T.}~\bibnamefont {Hao}}, \bibinfo {author} {\bibfnamefont {J.}~\bibnamefont
  {Rak}}, \bibinfo {author} {\bibfnamefont {J.}~\bibnamefont {Tavernier}},
  \bibinfo {author} {\bibfnamefont {D.~E.}\ \bibnamefont {Hill}}, \bibinfo
  {author} {\bibfnamefont {M.}~\bibnamefont {Vidal}}, \bibinfo {author}
  {\bibfnamefont {F.~P.}\ \bibnamefont {Roth}}, \ and\ \bibinfo {author}
  {\bibfnamefont {M.~A.}\ \bibnamefont {Calderwood}},\ }\bibfield  {title}
  {\emph {\bibinfo {title} {{A reference map of the human binary protein
  interactome}},\ }}\href {\doibase 10.1038/s41586-020-2188-x} {\bibfield
  {journal} {\bibinfo  {journal} {Nature}\ }\textbf {\bibinfo {volume} {580}},\
  \bibinfo {pages} {402} (\bibinfo {year} {2020})}\BibitemShut {NoStop}%
\bibitem [{\citenamefont {Yu}\ \emph {et~al.}(2008)\citenamefont {Yu},
  \citenamefont {Braun}, \citenamefont {Y{\i}ld{\i}r{\i}m}, \citenamefont {Lemmens},
  \citenamefont {Venkatesan}, \citenamefont {Sahalie}, \citenamefont
  {Hirozane-Kishikawa}, \citenamefont {Gebreab}, \citenamefont {Li},
  \citenamefont {Simonis}, \citenamefont {Hao}, \citenamefont {Rual},
  \citenamefont {Dricot}, \citenamefont {Vazquez}, \citenamefont {Murray},
  \citenamefont {Simon}, \citenamefont {Tardivo}, \citenamefont {Tam},
  \citenamefont {Svrzikapa}, \citenamefont {Fan}, \citenamefont {de~Smet},
  \citenamefont {Motyl}, \citenamefont {Hudson}, \citenamefont {Park},
  \citenamefont {Xin}, \citenamefont {Cusick}, \citenamefont {Moore},
  \citenamefont {Boone}, \citenamefont {Snyder}, \citenamefont {Roth},
  \citenamefont {Barab\'{a}si}, \citenamefont {Tavernier}, \citenamefont {Hill},\ and\ \citenamefont {Vidal}}]{Yu2008}%
  \BibitemOpen
  \bibfield  {author} {\bibinfo {author} {\bibfnamefont {H.}~\bibnamefont
  {Yu}}, \bibinfo {author} {\bibfnamefont {P.}~\bibnamefont {Braun}}, \bibinfo
  {author} {\bibfnamefont {M.~A.}\ \bibnamefont {Y{\i}ld{\i}r{\i}m}}, \bibinfo
  {author} {\bibfnamefont {I.}~\bibnamefont {Lemmens}}, \bibinfo {author}
  {\bibfnamefont {K.}~\bibnamefont {Venkatesan}}, \bibinfo {author}
  {\bibfnamefont {J.}~\bibnamefont {Sahalie}}, \bibinfo {author} {\bibfnamefont
  {T.}~\bibnamefont {Hirozane-Kishikawa}}, \bibinfo {author} {\bibfnamefont
  {F.}~\bibnamefont {Gebreab}}, \bibinfo {author} {\bibfnamefont
  {N.}~\bibnamefont {Li}}, \bibinfo {author} {\bibfnamefont {N.}~\bibnamefont
  {Simonis}}, \bibinfo {author} {\bibfnamefont {T.}~\bibnamefont {Hao}},
  \bibinfo {author} {\bibfnamefont {J.-F.}\ \bibnamefont {Rual}}, \bibinfo
  {author} {\bibfnamefont {A.}~\bibnamefont {Dricot}}, \bibinfo {author}
  {\bibfnamefont {A.}~\bibnamefont {Vazquez}}, \bibinfo {author} {\bibfnamefont
  {R.~R.}\ \bibnamefont {Murray}}, \bibinfo {author} {\bibfnamefont
  {C.}~\bibnamefont {Simon}}, \bibinfo {author} {\bibfnamefont
  {L.}~\bibnamefont {Tardivo}}, \bibinfo {author} {\bibfnamefont
  {S.}~\bibnamefont {Tam}}, \bibinfo {author} {\bibfnamefont {N.}~\bibnamefont
  {Svrzikapa}}, \bibinfo {author} {\bibfnamefont {C.}~\bibnamefont {Fan}},
  \bibinfo {author} {\bibfnamefont {A.-S.}\ \bibnamefont {de~Smet}}, \bibinfo
  {author} {\bibfnamefont {A.}~\bibnamefont {Motyl}}, \bibinfo {author}
  {\bibfnamefont {M.~E.}\ \bibnamefont {Hudson}}, \bibinfo {author}
  {\bibfnamefont {J.}~\bibnamefont {Park}}, \bibinfo {author} {\bibfnamefont
  {X.}~\bibnamefont {Xin}}, \bibinfo {author} {\bibfnamefont {M.~E.}\
  \bibnamefont {Cusick}}, \bibinfo {author} {\bibfnamefont {T.}~\bibnamefont
  {Moore}}, \bibinfo {author} {\bibfnamefont {C.}~\bibnamefont {Boone}},
  \bibinfo {author} {\bibfnamefont {M.}~\bibnamefont {Snyder}}, \bibinfo
  {author} {\bibfnamefont {F.~P.}\ \bibnamefont {Roth}}, \bibinfo {author}
  {\bibfnamefont {A.-L.}\ \bibnamefont {Barab\'{a}si}}, \bibinfo {author}
  {\bibfnamefont {J.}~\bibnamefont {Tavernier}}, \bibinfo {author}
  {\bibfnamefont {D.~E.}\ \bibnamefont {Hill}}, \ and\ \bibinfo {author}
  {\bibfnamefont {M.}~\bibnamefont {Vidal}},\ }\bibfield  {title} {\emph
  {\bibinfo {title} {{High-Quality Binary Protein Interaction Map of the Yeast
  Interactome Network}},\ }}\href {\doibase 10.1126/science.1158684} {\bibfield
   {journal} {\bibinfo  {journal} {Science}\ }\textbf {\bibinfo {volume}
  {322}},\ \bibinfo {pages} {104} (\bibinfo {year} {2008})}\BibitemShut
  {NoStop}%
\bibitem [{\citenamefont {Dunne}\ \emph {et~al.}(2014)\citenamefont {Dunne},
  \citenamefont {Labandeira},\ and\ \citenamefont
  {Williams}}]{dunne2014highly}%
  \BibitemOpen
  \bibfield  {author} {\bibinfo {author} {\bibfnamefont {J.~A.}\ \bibnamefont
  {Dunne}}, \bibinfo {author} {\bibfnamefont {C.~C.}\ \bibnamefont
  {Labandeira}}, \ and\ \bibinfo {author} {\bibfnamefont {R.~J.}\ \bibnamefont
  {Williams}},\ }\bibfield  {title} {\emph {\bibinfo {title} {{Highly resolved
  early Eocene food webs show development of modern trophic structure after the
  end-Cretaceous extinction}},\ }}\href {\doibase 10.1098/rspb.2013.3280}
  {\bibfield  {journal} {\bibinfo  {journal} {Proc. R. Soc. B Biol. Sci.}\
  }\textbf {\bibinfo {volume} {281}},\ \bibinfo {pages} {20133280} (\bibinfo
  {year} {2014})}\BibitemShut {NoStop}%
\bibitem [{\citenamefont {Kunegis}(2013)}]{kunegis2013konect}%
  \BibitemOpen
  \bibfield  {author} {\bibinfo {author} {\bibfnamefont {J.}~\bibnamefont
  {Kunegis}},\ }in\ \href@noop {} {\emph {\bibinfo {booktitle} {WWW 2013
  Companion - Proc. 22nd Int. Conf. World Wide Web}}}\ (\bibinfo {year}
  {2013})\BibitemShut {NoStop}%
\bibitem [{\citenamefont {Perozzi}\ \emph {et~al.}(2014)\citenamefont
  {Perozzi}, \citenamefont {Al-Rfou},\ and\ \citenamefont
  {Skiena}}]{perozzi2014deep}%
  \BibitemOpen
  \bibfield  {author} {\bibinfo {author} {\bibfnamefont {B.}~\bibnamefont
  {Perozzi}}, \bibinfo {author} {\bibfnamefont {R.}~\bibnamefont {Al-Rfou}}, \
  and\ \bibinfo {author} {\bibfnamefont {S.}~\bibnamefont {Skiena}},\ }in\
  \href {\doibase 10.1145/2623330.2623732} {\emph {\bibinfo {booktitle} {Proc.
  20th ACM SIGKDD Int. Conf. Knowl. Discov. data Min.}}}\ (\bibinfo
  {publisher} {ACM},\ \bibinfo {address} {New York, NY, USA},\ \bibinfo {year}
  {2014})\ pp.\ \bibinfo {pages} {701--710}\BibitemShut {NoStop}%
\bibitem [{\citenamefont {Zhou}\ \emph {et~al.}(2009)\citenamefont {Zhou},
  \citenamefont {L{\"{u}}},\ and\ \citenamefont {Zhang}}]{Zhou2009predicting}%
  \BibitemOpen
  \bibfield  {author} {\bibinfo {author} {\bibfnamefont {T.}~\bibnamefont
  {Zhou}}, \bibinfo {author} {\bibfnamefont {L.}~\bibnamefont {L{\"{u}}}}, \
  and\ \bibinfo {author} {\bibfnamefont {Y.-C.}\ \bibnamefont {Zhang}},\
  }\bibfield  {title} {\emph {\bibinfo {title} {{Predicting missing links via
  local information}},\ }}\href {\doibase 10.1140/epjb/e2009-00335-8}
  {\bibfield  {journal} {\bibinfo  {journal} {Eur. Phys. J. B}\ }\textbf
  {\bibinfo {volume} {71}},\ \bibinfo {pages} {623} (\bibinfo {year}
  {2009})}\BibitemShut {NoStop}%
\bibitem [{\citenamefont {Adamic}\ and\ \citenamefont
  {Adar}(2003)}]{Adamic2003friends}%
  \BibitemOpen
  \bibfield  {author} {\bibinfo {author} {\bibfnamefont {L.~A.}\ \bibnamefont
  {Adamic}}\ and\ \bibinfo {author} {\bibfnamefont {E.}~\bibnamefont {Adar}},\
  }\bibfield  {title} {\emph {\bibinfo {title} {{Friends and neighbors on the
  Web}},\ }}\href {\doibase 10.1016/S0378-8733(03)00009-1} {\bibfield
  {journal} {\bibinfo  {journal} {Soc. Networks}\ }\textbf {\bibinfo {volume}
  {25}},\ \bibinfo {pages} {211} (\bibinfo {year} {2003})}\BibitemShut
  {NoStop}%
\bibitem [{\citenamefont {Barab\'{a}si}\ and\ \citenamefont
  {Albert}(1999)}]{Barabasi1999}%
  \BibitemOpen
  \bibfield  {author} {\bibinfo {author} {\bibfnamefont {A.-L.}\ \bibnamefont
  {Barab\'{a}si}}\ and\ \bibinfo {author} {\bibfnamefont {R.}~\bibnamefont
  {Albert}},\ }\bibfield  {title} {\emph {\bibinfo {title} {{Emergence of
  Scaling in Random Networks}},\ }}\href {\doibase
  10.1126/science.286.5439.509} {\bibfield  {journal} {\bibinfo  {journal}
  {Science}\ }\textbf {\bibinfo {volume} {286}},\ \bibinfo {pages} {509}
  (\bibinfo {year} {1999})}\BibitemShut {NoStop}%
\bibitem [{\citenamefont {L{\"{u}}}\ \emph {et~al.}(2015)\citenamefont
  {L{\"{u}}}, \citenamefont {Pan}, \citenamefont {Zhou}, \citenamefont
  {Zhang},\ and\ \citenamefont {Stanley}}]{lu2015toward}%
  \BibitemOpen
  \bibfield  {author} {\bibinfo {author} {\bibfnamefont {L.}~\bibnamefont
  {L{\"{u}}}}, \bibinfo {author} {\bibfnamefont {L.}~\bibnamefont {Pan}},
  \bibinfo {author} {\bibfnamefont {T.}~\bibnamefont {Zhou}}, \bibinfo {author}
  {\bibfnamefont {Y.-C.}\ \bibnamefont {Zhang}}, \ and\ \bibinfo {author}
  {\bibfnamefont {H.~E.}\ \bibnamefont {Stanley}},\ }\bibfield  {title} {\emph
  {\bibinfo {title} {{Toward link predictability of complex networks}},\
  }}\href {\doibase 10.1073/pnas.1424644112} {\bibfield  {journal} {\bibinfo
  {journal} {Proc. Natl. Acad. Sci.}\ }\textbf {\bibinfo {volume} {112}},\
  \bibinfo {pages} {2325} (\bibinfo {year} {2015})}\BibitemShut {NoStop}%
\bibitem [{\citenamefont {Katz}(1953)}]{Katz1953new}%
  \BibitemOpen
  \bibfield  {author} {\bibinfo {author} {\bibfnamefont {L.}~\bibnamefont
  {Katz}},\ }\bibfield  {title} {\emph {\bibinfo {title} {{A new status index
  derived from sociometric analysis}},\ }}\href {\doibase 10.1007/BF02289026}
  {\bibfield  {journal} {\bibinfo  {journal} {Psychometrika}\ }\textbf
  {\bibinfo {volume} {18}},\ \bibinfo {pages} {39} (\bibinfo {year}
  {1953})}\BibitemShut {NoStop}%
\bibitem [{\citenamefont {Cannistraci}\ \emph
  {et~al.}(2013{\natexlab{b}})\citenamefont {Cannistraci}, \citenamefont
  {Alanis-Lobato},\ and\ \citenamefont {Ravasi}}]{Cannistraci2013b}%
  \BibitemOpen
  \bibfield  {author} {\bibinfo {author} {\bibfnamefont {C.~V.}\ \bibnamefont
  {Cannistraci}}, \bibinfo {author} {\bibfnamefont {G.}~\bibnamefont
  {Alanis-Lobato}}, \ and\ \bibinfo {author} {\bibfnamefont {T.}~\bibnamefont
  {Ravasi}},\ }\bibfield  {title} {\emph {\bibinfo {title} {{From
  link-prediction in brain connectomes and protein interactomes to the
  local-community-paradigm in complex networks}},\ }}\href {\doibase
  10.1038/srep01613} {\bibfield  {journal} {\bibinfo  {journal} {Sci. Rep.}\
  }\textbf {\bibinfo {volume} {3}},\ \bibinfo {pages} {1613} (\bibinfo {year}
  {2013}{\natexlab{b}})}\BibitemShut {NoStop}%
\bibitem [{\citenamefont {Jaccard}(1901)}]{Jaccard1901}%
  \BibitemOpen
  \bibfield  {author} {\bibinfo {author} {\bibfnamefont {P.}~\bibnamefont
  {Jaccard}},\ }\bibfield  {title} {\emph {\bibinfo {title} {{{\'{E}}tude
  comparative de la distribution florale dans une portion des Alpes et des
  Jura}},\ }}\href@noop {} {\bibfield  {journal} {\bibinfo  {journal} {Bull.
  del la Soci{\'{e}}t{\'{e}} Vaudoise des Sci. Nat.}\ }\textbf {\bibinfo
  {volume} {37}},\ \bibinfo {pages} {547} (\bibinfo {year} {1901})}\BibitemShut
  {NoStop}%
\bibitem [{\citenamefont {Liben-Nowell}\ and\ \citenamefont
  {Kleinberg}(2003)}]{Liben2003link}%
  \BibitemOpen
  \bibfield  {author} {\bibinfo {author} {\bibfnamefont {D.}~\bibnamefont
  {Liben-Nowell}}\ and\ \bibinfo {author} {\bibfnamefont {J.}~\bibnamefont
  {Kleinberg}},\ }in\ \href {\doibase 10.1145/956863.956972} {\emph {\bibinfo
  {booktitle} {Proc. twelfth Int. Conf. Inf. Knowl. Manag.}}}\ (\bibinfo
  {publisher} {ACM},\ \bibinfo {address} {New York, NY, USA},\ \bibinfo {year}
  {2003})\ pp.\ \bibinfo {pages} {556--559}\BibitemShut {NoStop}%
\bibitem [{\citenamefont {Ghiassian}\ \emph {et~al.}(2015)\citenamefont
  {Ghiassian}, \citenamefont {Menche},\ and\ \citenamefont
  {Barab{\'{a}}si}}]{ghiassian2015disease}%
  \BibitemOpen
  \bibfield  {author} {\bibinfo {author} {\bibfnamefont {S.~D.}\ \bibnamefont
  {Ghiassian}}, \bibinfo {author} {\bibfnamefont {J.}~\bibnamefont {Menche}}, \
  and\ \bibinfo {author} {\bibfnamefont {A.-L.}\ \bibnamefont
  {Barab{\'{a}}si}},\ }\bibfield  {title} {\emph {\bibinfo {title} {{A DIseAse
  MOdule Detection (DIAMOnD) Algorithm Derived from a Systematic Analysis of
  Connectivity Patterns of Disease Proteins in the Human Interactome}},\
  }}\href {\doibase 10.1371/journal.pcbi.1004120} {\bibfield  {journal}
  {\bibinfo  {journal} {PLOS Comput. Biol.}\ }\textbf {\bibinfo {volume}
  {11}},\ \bibinfo {pages} {e1004120} (\bibinfo {year} {2015})}\BibitemShut
  {NoStop}%
\bibitem [{\citenamefont {Ahn}\ \emph {et~al.}(2010)\citenamefont {Ahn},
  \citenamefont {Bagrow},\ and\ \citenamefont {Lehmann}}]{ahn2010link}%
  \BibitemOpen
  \bibfield  {author} {\bibinfo {author} {\bibfnamefont {Y.-Y.}\ \bibnamefont
  {Ahn}}, \bibinfo {author} {\bibfnamefont {J.~P.}\ \bibnamefont {Bagrow}}, \
  and\ \bibinfo {author} {\bibfnamefont {S.}~\bibnamefont {Lehmann}},\
  }\bibfield  {title} {\emph {\bibinfo {title} {{Link communities reveal
  multiscale complexity in networks}},\ }}\href {\doibase 10.1038/nature09182}
  {\bibfield  {journal} {\bibinfo  {journal} {Nature}\ }\textbf {\bibinfo
  {volume} {466}},\ \bibinfo {pages} {761} (\bibinfo {year} {2010})},\ \Eprint
  {http://arxiv.org/abs/0903.3178} {arXiv:0903.3178} \BibitemShut {NoStop}%
\bibitem [{\citenamefont {Mahmoud}\ \emph {et~al.}(2014)\citenamefont
  {Mahmoud}, \citenamefont {Masulli}, \citenamefont {Rovetta},\ and\
  \citenamefont {Russo}}]{mahmoud2014comm}%
  \BibitemOpen
  \bibfield  {author} {\bibinfo {author} {\bibfnamefont {H.}~\bibnamefont
  {Mahmoud}}, \bibinfo {author} {\bibfnamefont {F.}~\bibnamefont {Masulli}},
  \bibinfo {author} {\bibfnamefont {S.}~\bibnamefont {Rovetta}}, \ and\
  \bibinfo {author} {\bibfnamefont {G.}~\bibnamefont {Russo}},\ }in\ \href
  {\doibase 10.1007/978-3-319-09042-9_5} {\emph {\bibinfo {booktitle} {Lect.
  Notes Comput. Sci.}}},\ Vol.\ \bibinfo {volume} {8452 LNBI}\ (\bibinfo
  {publisher} {Springer Verlag},\ \bibinfo {year} {2014})\ pp.\ \bibinfo
  {pages} {62--75}\BibitemShut {NoStop}%
\bibitem [{\citenamefont {Girvan}\ and\ \citenamefont
  {Newman}(2002)}]{Girvan2002a}%
  \BibitemOpen
  \bibfield  {author} {\bibinfo {author} {\bibfnamefont {M.}~\bibnamefont
  {Girvan}}\ and\ \bibinfo {author} {\bibfnamefont {M.~E.~J.}\ \bibnamefont
  {Newman}},\ }\bibfield  {title} {\emph {\bibinfo {title} {{Community
  structure in social and biological networks}},\ }}\href {\doibase
  10.1073/pnas.122653799} {\bibfield  {journal} {\bibinfo  {journal} {Proc.
  Natl. Acad. Sci.}\ }\textbf {\bibinfo {volume} {99}},\ \bibinfo {pages}
  {7821} (\bibinfo {year} {2002})}\BibitemShut {NoStop}%
\bibitem [{\citenamefont {Fortunato}(2010)}]{fortunato2010community}%
  \BibitemOpen
  \bibfield  {author} {\bibinfo {author} {\bibfnamefont {S.}~\bibnamefont
  {Fortunato}},\ }\bibfield  {title} {\emph {\bibinfo {title} {{Community
  detection in graphs}},\ }}\href {\doibase 10.1016/j.physrep.2009.11.002}
  {\bibfield  {journal} {\bibinfo  {journal} {Phys. Rep.}\ }\textbf {\bibinfo
  {volume} {486}},\ \bibinfo {pages} {75} (\bibinfo {year} {2010})}\BibitemShut
  {NoStop}%
\bibitem [{\citenamefont {Menche}\ \emph {et~al.}(2015)\citenamefont {Menche},
  \citenamefont {Sharma}, \citenamefont {Kitsak}, \citenamefont {Ghiassian},
  \citenamefont {Vidal}, \citenamefont {Loscalzo},\ and\ \citenamefont
  {Barabasi}}]{menche2015uncovering}%
  \BibitemOpen
  \bibfield  {author} {\bibinfo {author} {\bibfnamefont {J.}~\bibnamefont
  {Menche}}, \bibinfo {author} {\bibfnamefont {A.}~\bibnamefont {Sharma}},
  \bibinfo {author} {\bibfnamefont {M.}~\bibnamefont {Kitsak}}, \bibinfo
  {author} {\bibfnamefont {S.~D.}\ \bibnamefont {Ghiassian}}, \bibinfo {author}
  {\bibfnamefont {M.}~\bibnamefont {Vidal}}, \bibinfo {author} {\bibfnamefont
  {J.}~\bibnamefont {Loscalzo}}, \ and\ \bibinfo {author} {\bibfnamefont
  {A.-L.}\ \bibnamefont {Barab\'{a}si}},\ }\bibfield  {title} {\emph {\bibinfo
  {title} {{Uncovering disease-disease relationships through the incomplete
  interactome}},\ }}\href {\doibase 10.1126/science.1257601} {\bibfield
  {journal} {\bibinfo  {journal} {Science}\ }\textbf {\bibinfo {volume}
  {347}},\ \bibinfo {pages} {1257601} (\bibinfo {year} {2015})}\BibitemShut
  {NoStop}%
\bibitem [{\citenamefont {Sonawane}\ \emph {et~al.}(2019)\citenamefont
  {Sonawane}, \citenamefont {Weiss}, \citenamefont {Glass},\ and\ \citenamefont
  {Sharma}}]{Sonawane2019}%
  \BibitemOpen
  \bibfield  {author} {\bibinfo {author} {\bibfnamefont {A.~R.}\ \bibnamefont
  {Sonawane}}, \bibinfo {author} {\bibfnamefont {S.~T.}\ \bibnamefont {Weiss}},
  \bibinfo {author} {\bibfnamefont {K.}~\bibnamefont {Glass}}, \ and\ \bibinfo
  {author} {\bibfnamefont {A.}~\bibnamefont {Sharma}},\ }\bibfield  {title}
  {\emph {\bibinfo {title} {{Network Medicine in the Age of Biomedical Big
  Data}},\ }}\href {\doibase 10.3389/fgene.2019.00294} {\bibfield  {journal}
  {\bibinfo  {journal} {Front. Genet.}\ }\textbf {\bibinfo {volume} {10}}
  (\bibinfo {year} {2019}),\ 10.3389/fgene.2019.00294},\ \Eprint
  {http://arxiv.org/abs/1903.05449} {arXiv:1903.05449} \BibitemShut {NoStop}%
\bibitem [{\citenamefont {Speer}\ \emph {et~al.}(2017)\citenamefont {Speer},
  \citenamefont {Chin},\ and\ \citenamefont {Havasi}}]{speer2017conceptnet}%
  \BibitemOpen
  \bibfield  {author} {\bibinfo {author} {\bibfnamefont {R.}~\bibnamefont
  {Speer}}, \bibinfo {author} {\bibfnamefont {J.}~\bibnamefont {Chin}}, \ and\
  \bibinfo {author} {\bibfnamefont {C.}~\bibnamefont {Havasi}},\ }\href
  {http://aaai.org/ocs/index.php/AAAI/AAAI17/paper/view/14972} {\bibfield
  {title} {\emph {\bibinfo {title} {ConceptNet 5.5: An Open Multilingual Graph
  of General Knowledge},\ }}} (\bibinfo {year} {2017})\BibitemShut {NoStop}%
\bibitem [{\citenamefont {Stark}\ \emph {et~al.}(1984)\citenamefont {Stark},
  \citenamefont {Breitkreutz}, \citenamefont {Reguly}, \citenamefont {Boucher},
  \citenamefont {Breitkreutz},\ and\ \citenamefont {Tyers}}]{Stark2006}%
  \BibitemOpen
  \bibfield  {author} {\bibinfo {author} {\bibfnamefont {C.}~\bibnamefont
  {Stark}}, \bibinfo {author} {\bibfnamefont {B.-J.}\ \bibnamefont
  {Breitkreutz}}, \bibinfo {author} {\bibfnamefont {T.}~\bibnamefont {Reguly}},
  \bibinfo {author} {\bibfnamefont {L.}~\bibnamefont {Boucher}}, \bibinfo
  {author} {\bibfnamefont {A.}~\bibnamefont {Breitkreutz}}, \ and\ \bibinfo
  {author} {\bibfnamefont {M.}~\bibnamefont {Tyers}},\ }\bibfield  {title}
  {\emph {\bibinfo {title} {{BioGRID: a general repository for interaction
  datasets.}}\ }}\href {\doibase 10.1093/nar/gkj109} {\bibfield  {journal}
  {\bibinfo  {journal} {Nucleic Acids Res.}\ }\textbf {\bibinfo {volume}
  {34}},\ \bibinfo {pages} {D535} (\bibinfo {year} {1984})}\BibitemShut
  {NoStop}%
\bibitem [{cod(2020)}]{codeHLembedder}%
  \BibitemOpen
  \href@noop {} {\bibfield  {title} {\emph {\bibinfo {title} {HyperLink
  embedder},\ }}} (\bibinfo {year} {2020}),\ \bibinfo {note}
  {\url{https://bitbucket.org/dk-lab/2020_code_hyperlink}}\BibitemShut
  {NoStop}%
\bibitem [{\citenamefont {Krioukov}(1984)}]{krioukov2016clustering}%
  \BibitemOpen
  \bibfield  {author} {\bibinfo {author} {\bibfnamefont {D.}~\bibnamefont
  {Krioukov}},\ }\bibfield  {title} {\emph {\bibinfo {title} {{Clustering
  Implies Geometry in Networks}},\ }}\href {\doibase
  10.1103/PhysRevLett.116.208302} {\bibfield  {journal} {\bibinfo  {journal}
  {Phys. Rev. Lett.}\ }\textbf {\bibinfo {volume} {116}} (\bibinfo {year}
  {1984}),\ 10.1103/PhysRevLett.116.208302}\BibitemShut {NoStop}%
\end{thebibliography}

%

\end{document}